\documentclass[sigconf]{acmart}
\usepackage{booktabs} % For formal tables
\usepackage{epsfig}
\usepackage{amssymb}
\usepackage{amsmath}
\usepackage{tabularx}
\usepackage{tabularx}
\usepackage{tabularx,ragged2e}
\newcolumntype{C}{>{\Centering\arraybackslash}X} % centered "X" column
\usepackage{amsfonts}
\usepackage[utf8]{inputenc}
\usepackage[T1]{fontenc}
\usepackage{tabu}
\usepackage{hyperref}
\usepackage{multirow}
\usepackage{makecell}
\usepackage{ctable}
\usepackage{capt-of}
\usepackage{varwidth}
\usepackage{balance}

\copyrightyear{2018}
\acmYear{2018}
\setcopyright{acmlicensed}
\acmConference[MM '18]{2018 ACM Multimedia Conference}{October 22--26, 2018}{Seoul, Republic of Korea}
\acmPrice{15.00}
\acmDOI{10.1145/3240508.3241913}
\acmISBN{978-1-4503-5665-7/18/10}

\hypersetup{draft}
\begin{document}
\title{Cross-Modal Health State Estimation}

%  via Multi-Modal Data
% \titlenote{Produces the permission block, and
% \subtitle{Extended Abstract}
% \subtitlenote{The full version of the author's guide is available as
%   \texttt{acmart.pdf} document}

% \author{Anonymous}
% % \authornote{Dr.~Trovato insisted his name be first.}
% % \orcid{1234-5678-9012}
% \affiliation{%
%   \institution{}
% %   \streetaddress{P.O. Box 1212}
%   \city{} 
%   \state{} 
% %   \postcode{43017-6221}
% }
% \email{}

\author{Nitish Nag}
% \authornote{Dr.~Trovato insisted his name be first.}
\affiliation{%
  \institution{University of California, Irvine}
  \city{Irvine} 
  \state{California} 
}
\email{nagn@uci.edu}

\author{Vaibhav Pandey}
% \authornote{Dr.~Trovato insisted his name be first.}
\affiliation{%
  \institution{University of California, Irvine}
  \city{Irvine} 
  \state{California} 
}
\email{vaibhap1@uci.edu}

\author{Preston J. Putzel}
% \authornote{The secretary disavows any knowledge of this author's actions.}
\affiliation{%
  \institution{University of California, Irvine}
  \city{Irvine} 
  \state{California} 
}
\email{pputzel@uci.edu}

\author{Hari Bhimaraju}
% \authornote{The secretary disavows any knowledge of this author's actions.}
\affiliation{%
  \institution{University of California, Irvine}
  \city{Irvine} 
  \state{California} 
}
\email{haribhimaraju@gmail.com}

\author{Srikanth Krishnan}
% \authornote{Dr.~Trovato insisted his name be first.}
\affiliation{%
  \institution{University of California, Los Angeles}
  \city{Los Angeles} 
  \state{California} 
}
\email{srikanthkrishnan@mednet.ucla.edu}

\author{Ramesh Jain}
% \authornote{The secretary disavows any knowledge of this author's actions.}
\affiliation{%
  \institution{University of California, Irvine}
  \city{Irvine} 
  \state{California} 
}
\email{jain@ics.uci.edu}

\begin{abstract}
Individuals create and consume more diverse data about themselves today than any time in history. Sources of this data include wearable devices, images, social media, geo-spatial information and more. A tremendous opportunity rests within cross-modal data analysis that leverages existing domain knowledge methods to understand and guide human health. Especially in chronic diseases, current medical practice uses a combination of sparse hospital based biological metrics (blood tests, expensive imaging, etc.) to understand the evolving health status of an individual. Future health systems must integrate data created at the individual level to better understand health status perpetually, especially in a cybernetic framework. In this work we fuse multiple user created and open source data streams along with established biomedical domain knowledge to give two types of quantitative state estimates of cardiovascular health. First, we use wearable devices to calculate cardiorespiratory fitness (CRF), a known quantitative leading predictor of heart disease which is not routinely collected in clinical settings. Second, we estimate inherent genetic traits, living environmental risks, circadian rhythm, and biological metrics from a diverse dataset. Our experimental results on 24 subjects demonstrate how multi-modal data can provide personalized health insight. Understanding the dynamic nature of health status will pave the way for better health based recommendation engines, better clinical decision making and positive lifestyle changes.
\end{abstract}

\ccsdesc[500]{Information systems~Multimedia information systems}
\ccsdesc[500]{Information systems~Multimedia and multimodal retrieval}
\ccsdesc[500]{Social and professional topics~Personal health records}
\ccsdesc[500]{Applied computing~Systems biology}
\ccsdesc[500]{Applied computing~Consumer health}
\ccsdesc[500]{Applied computing~Health care information systems}
\ccsdesc[500]{Applied computing~Health informatics}
\ccsdesc[500]{Applied computing~Bioinformatics}
\ccsdesc[500]{Computer systems organization~Embedded and cyber-physical systems}
\ccsdesc[300]{General and reference~Estimation}
\ccsdesc[300]{Networks~Cyber-physical networks}

\keywords{Personal Health Navigation; Cross-Modal Data; ; Health Situation; Cybernetic Health; Multimedia; Wearables; Health Informatics}

% \acmBadgeR{artifacts_available}

%% Used in some conference proceedings e.g. sigplan and sigchi
%   \caption{This is a teaser}

\maketitle

\section{Introduction}

"To live effectively is to live with adequate information."

\hspace*{\fill} - Norbert Weiner, 1950
\newline

\begin{figure}
\small
\centering
\includegraphics[width=1 \columnwidth]{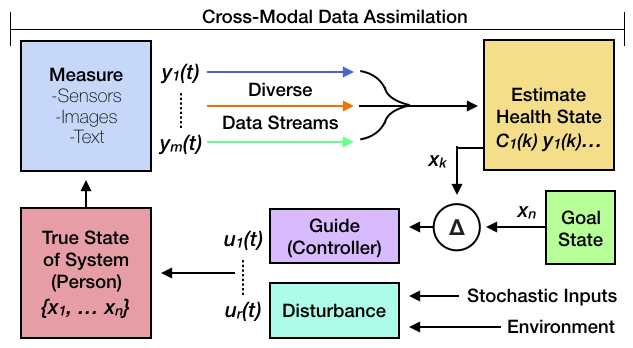}
\caption{Cross-modal measurements are essential for state estimation in cybernetic feedback systems. This estimation impacts the eventual guidance from the controller (physician or automated system) to help reach a goal state.}~\label{fig:mega}
\vspace{-5mm}
\end{figure}

A century ago, the largest contributor to mortality and morbidity was infectious disease. Infection is an episodic problem, where a categorical diagnosis (i.e. malaria) is made after a patient feels unwell and arrives at a medical care facility (where the data is gathered to confirm the disease). Treatment is usually prescribed based on evidence based rules to solve the problem, and the patient is not monitored anymore.  Globally, chronic diseases have emerged as the 21st century major contributor to health burden. When compared to infectious disease, there are fundamental differences. There is no single event that leads to the disease, rather slow changes in the operating function of the body. If we are to apply some type of control input to keep people on a healthy trajectory via feedback loops in cybernetic systems, we must be able to continuously estimate this state. Hence there is a clear distinction between the classification versus quantified estimation problem in health. Due to this reason, there is compelling need for progress in health state estimation.

\begin{figure}
\small
\centering
\includegraphics[width=.9 \columnwidth]{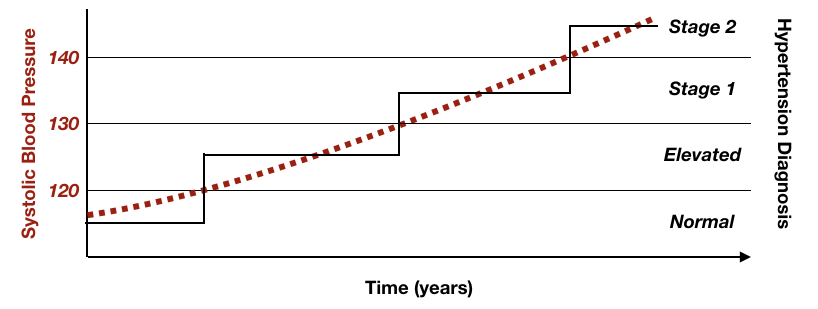}
\caption{True physiological average blood pressure (red dashed) vs. clinical assessment (black solid) show the comparison between health state assessments that are quantitative vs. categorical.}~\label{fig:htn}
\vspace{-8mm}
\end{figure}

To illustrate this need we describe the situation of hypertension (high blood pressure). First, patients are unable to feel the disease as it slowly builds up over time, and thus do not even know they are being affected. Second, the diseases stem from both daily actions and environmental exposures, not just a single source. Third, these diseases are not truly categorical in nature, but are rather declines in organ function over time. In the example of hypertension, clinical practice uses cutoff thresholds to decide when to change the labeled blood pressure status of an individual, when in reality, the average pressure is increasing over time as shown in Figure \ref{fig:htn}. Ultimately, individuals, clinicians, and in general cybernetic systems (Figure \ref{fig:mega}), make decisions based on the method of determining health state. What we measure is what we control.

Individuals create diverse data streams about themselves on a daily basis. Much of this data can be leveraged to provide perpetual insight into the health of individuals. Fitness devices and wearable sensors, ambient sensors, images, video, audio, digital human computer interactions, and IoT devices provide a plethora of data that is routinely collected, but so far has been difficult to use for common real world health applications. Because people are unable to feel their health change over time from the multitude of factors affecting them, we need to develop methods to quantify and report health status using continuously collected multi-modal data sources. If we are able to track changes before permanent organ dysfunction, we may be able to correct course and prevent or delay onset of chronic diseases. Finally, medical practice needs to shift from using episodic categorical definitions of health status, to a continuous quantitative measurement. 

With the rapidly increasing availability of low cost sensors in the last decade, there has been an explosion in the amount of continuously collected multi-modal data. This is especially relevant in field of health with the advent of wearable, IoT, and ambient sensors. Many of these low cost sensors such as accelerometers, light sensors, microphones, heart rate monitors, and barometers, to name a few, produce continuous streams of data usable in a wide range of scenarios. While this allows measurement of the user state continuously, the downside of these sensors is that the measurements are either very noisy, produce information overload, produce non-actionable metrics, or are not directly related to the attribute we want to measure. These limitations have so far proven to be a barrier in using low cost sensors for real world decision making for health. 

Furthermore, enabling data assimilation will require intelligible understanding of how sensor information relates to the health status. Existing medical and biological scientific domain knowledge must be used to guide the data assimilation and conversion matrices from signal to state. Intelligibility is also an important attribute for developing health estimation systems. Knowing \textit{why} an individual has a certain health status will be paramount to explanation, recommendation, and treatment.  Data driven methods can be components in a large system, but will have difficult time explaining the reason for the classification. This is why other complimentary methods need to be used that take advantage of domain knowledge.

\textbf{To summarize, we believe that cross-modal health state estimation will be a fundamental centerpiece for the following needs of future:}
\begin{itemize}
\item \textbf{Early Detection:} Insight into health status changes in the prodromal state, where health state can be readily altered towards wellness.
\item \textbf{Continuous Monitoring:} Understanding and assisting the individual in all aspects of life, everyday.
\item \textbf{Quantitative Real-Time Assessment:} Shifting health assessment to a dynamic quantitative measurement rather than categories of normal versus abnormal.
\item \textbf{Reduced Cost:} Through trickle down technology, we anticipate more data types available through devices, reducing the barriers and cost for health assessment.
\end{itemize}

Clinical need for measuring cardiorespiratory fitness (CRF) is in high demand, but at the moment it is only captured in high need care through expensive lab tests. In 2013, the American Heart Association and the American College of Cardiology jointly released guidelines for the prevention and treatment of coronary artery disease stating CRF is a leading risk factor for cardiovascular disease, the most significant cause of death in humans. Flatly stated by the AHA, "It is currently the only major risk factor not routinely assessed in clinical practice" \cite{Ross2016ImportanceAssociation}. The reason for not measuring this value for patients is due to the burdensome cost in time, inconvenience, and resources to gather this data directly. We take this as motivation to see if we can use lower cost wearable devices to accomplish this task. We compare how different wearable devices can provide observability into our own bodies. CRF levels change throughout our lives from effects of our lifestyle. A more refined and accurate reflection of cardiovascular health state would take into account additional information like the environment, stress, and genetic background. We address this challenge through adding additional data sources such as images and geospatial sensors.

For the aforementioned reasons, we focus the scope of our work on cardiovascular health state estimation by studying the following research questions:
\begin{itemize}
\item \textbf{Research Question 1 (RQ1):} What is the quality of CRF estimation via the combination of multi-modal data and domain knowledge from different wearable devices?
\item \textbf{Research Question 2 (RQ2):} What total cardiovascular health information can we elucidate from images, wearables, surveys, social media, Internet of Things (IoT), and environmental sensors? How can we assimilate this data in a useful way for individuals and health providers?
\end{itemize}

At the time of writing, there have been no investigations we found about cardiovascular health state measurements in the mutlimedia research community. Broadly, the motivation for this work in the Brave New Ideas track is to open the frontier into personalized health state estimation from multi-modal data. Further rigorous research in this field will look into expanding to other health domains, improve quality metrics, tackle performance issues, and much more. We hope ultimately to create research opportunities that allow us to effectively be informed about our health throughout life.

\section{Related Work} 
Health state estimation and tracking has been an important field in medical literature and computer science. There has been a strong call by the medical science community to use continuous multi-modal data for tracking individual health \cite{topol-2,Nag2017a,Nag2017b,SamGambhir2018TowardHealth,Wild2005ComplementingEpidemiology}. Most modern metrics that are used to understand patient health were derived from longitudinal studies of large cohorts to see what led to morbidity and mortality. Outcomes of these studies were then retrospectively analyzed with linear regression to predict future outcomes for new patients. Modern epidemiology efforts are beginning to use modern data collection tools such as social multimedia and wearable devices \cite{Farseev2017TweetLearning}. These efforts include the United States Precision Medicine Initiative led by President Obama \cite{Francis2015AMedicine}, Mobile Sensor Data to Knowledge \cite{Kumar2017PervasiveMD2K}, and Alphabet's Verily division \cite{Dorsey2017VerilyBiomarkers}. These research efforts may take decades before we have data available for meaningful insight, as they largely depend upon outcomes of mortality before they become sufficiently powerful.

Within the field of cardiology, the Framingham study laid the foundation for most modern clinical guidelines by the American Heart Association (AHA) and American College of Cardiology \cite{Goff20142013Guidelines,Wilson2002OverweightRisk}. AHA has also called for the specific metric of CRF as the most powerful predictor of cardiovascular health that is not routinely measured (mostly due to cost of expensive and laborious lab testing) \cite{Ross2016ImportanceAssociation}. Technically speaking, activity which measures general movement patterns (such as through wearable accelerometers) is a different risk factor than aerobic exercise work capacity (which is CRF). CRF has a much stronger established relationship with true cardiovascular health \cite{Williams2001PhysicalMeta-Analysis}. Widespread use of standard wearable accelerometers that measure steps or higher semantic activities like walking, jogging, biking are indicative of activity only, hence the need for wearables that can give estimates of CRF.

Wearable devices have been used to estimate energy expenditure through various computational approaches such as deep learning \cite{Zhu2015UsingSensors}, knowledge based regression \cite{Lester2009ValidatedSensor}, and data filtering and segmentation techniques \cite{Albinali2010UsingEstimation}.  Energy expenditure provides insight into the total amount of activity performed by an individual, but does not provide maximal work output to estimate CRF. 

Computational research in CRF prediction began in the 1970's with the formulation of exercise stress scoring metrics based on then newly available chest strap based heart rate monitors \cite{Banister1980PlanningTraining., Borresen2009ThePerformance}. Recently, contextual understanding improved the performance of heart rate based CRF estimation, and were further refined by calibrating custom algorithmic parameters for a particular user \cite{Altini2015PersonalizedModels,Altini2016CardiorespiratorySensors}. Heart rate data has also been used to derive additional features, such as vagal tone (commonly referred to has heart rate variability) and respiratory rate, to provide regression analysis more features for prediction \cite{Smolander2008AWorkers}. Improvements in accelerometer based CRF prediction have been achieved through body placement optimization \cite{Parkka2007EstimatingLocations}. The only known research at this point that has attempted use of multimodal data for CRF prediction has been done by Firstbeat Corporation which uses both heart rate and speed information with a proprietary algorithm to filter periods of heart rate that are indicative of steady state metabolism \cite{Firstbeat2014AutomatedData}.

Multifactorial cardiovascular health risks have been investigated in many of the large epidemiologic studies such as the Framingham study. Conclusions from these large studies are used in current clinical practice through the AtheroSclerotic CardioVascular Diease (ASCVD) calculator \cite{Goff20142013Guidelines}. This pooled cohort algorithm was based on linear regression analysis for four separate cohorts of individuals female blacks, male blacks, female whites, male whites. Other than ethnicity and gender, they take into account age, systolic and diastolic blood pressure, cholesterol (Total, HDL, LDL), smoking history, diabetes (binary field: yes or no), medication history (hypertension, statin, aspirin only), and is only applicable for patients in the age range of 40-79. The limitations of this calculator include the requirement of invasive blood data and non-consumer based lab processing. No integrations of environment, lifestyle, social determinants, or biological parameters that test real world function such as CRF are used in any clinical setting at the moment. Individual parameters such as local air and noise pollution have established as risks, but are not shown in any relevant way to clinicians. Wearable devices are not used in any clinical setting for cardiovascular disease presently.

\section{Cybernetic Health State Estimation}

In the simplest terms, cybernetics is about setting goals and devising action sequences to accomplish and maintain those goals in the presence of noise and disturbances \cite{NorbertWiener1948Cybernetics:Machine}. This is enabled by the availability of sensors that can estimate the system state from observations to perpetually feed this information back to the system. This generates new control signals as required to move toward the desired goal or destination. Cybernetic in health has 4 main components: Measurement, Estimation, Guidance, Action as shown in figure \ref{fig:mega} \cite{Jain2018AHealth}. These four parts synthesize how we can produce a navigational system for improving health.

The mathematical model in classic systems theory states that:
\[ X[k+1] = A[k]X[k] + B[k]U[k]\]
\[ Y[k] = C[k]X[k] + D[k]U[k]\]

Where \textit{X}, \textit{U}, and \textit{Y} are the system true state, inputs, and measured output vectors respectively. \textit{A}, \textit{B}, \textit{C}, and \textit{D} are matrices that provide the appropriate transformation of these variables at a given time \textit{k}. Human health can be described by a state system, and the previous state and the inputs into the system play a role in determining health at time \textit{k+1}. Inputs into the human cybernetic system can be defined as anything which changes gene expression or physical actions in the body (from a molecular interactions to coarse movements). Thus a body is continuously exposed to these inputs which may or may not be within the controllability of an individual. 
The inputs beyond the control of an individual are referred to as external disturbance and the rest can be viewed as controllable inputs \textit{u}.
% Those inputs that are not within the control of an individual, we define as external disturbance. 
% Those that can be controlled, we call them controllable inputs \textit{u}. 
The true health state of an individual at a time \textit{k} is represented by \textit{X}, which is in reality difficult to obtain and always estimated. What we do get are the observable output variables. The state estimation challenge is in interpreting the observables to understand the underlying true state. If we solely focus on this, the challenge of state estimation is represented in the matrix \textit{C}, with observables as \textit{Y} and our unknown state as \textit{X}. 

Estimating health impairment in individuals who seem to be healthy is inherently difficult due to: 1)poor sensing ability of developing adverse outcomes with current clinical methods and 2)the long lead time to developing full blown chronic disease. By the time current clinical measurements such as cholesterol, blood pressure, or glucose metabolism are beyond the normal range, the user has already been in a dysfunctional health state for quite some time. Capturing the change in health state earlier (before true dysfunction begins) is paramount to keeping people healthy and preventing them from slipping into a diseased state. Clinical researchers refer to this as the prodromal state. Multimedia work in understanding, vision, classifiers, intent, and sentiment analysis can greatly expand the capability for higher resolution understanding of an individuals health state. In our following experimental work, we focus on this specific aspect in the domain of cardiovascular health.

\section{Experimental Approach}

\begin{figure*}
\small
\centering
\includegraphics[width=1.8 \columnwidth]{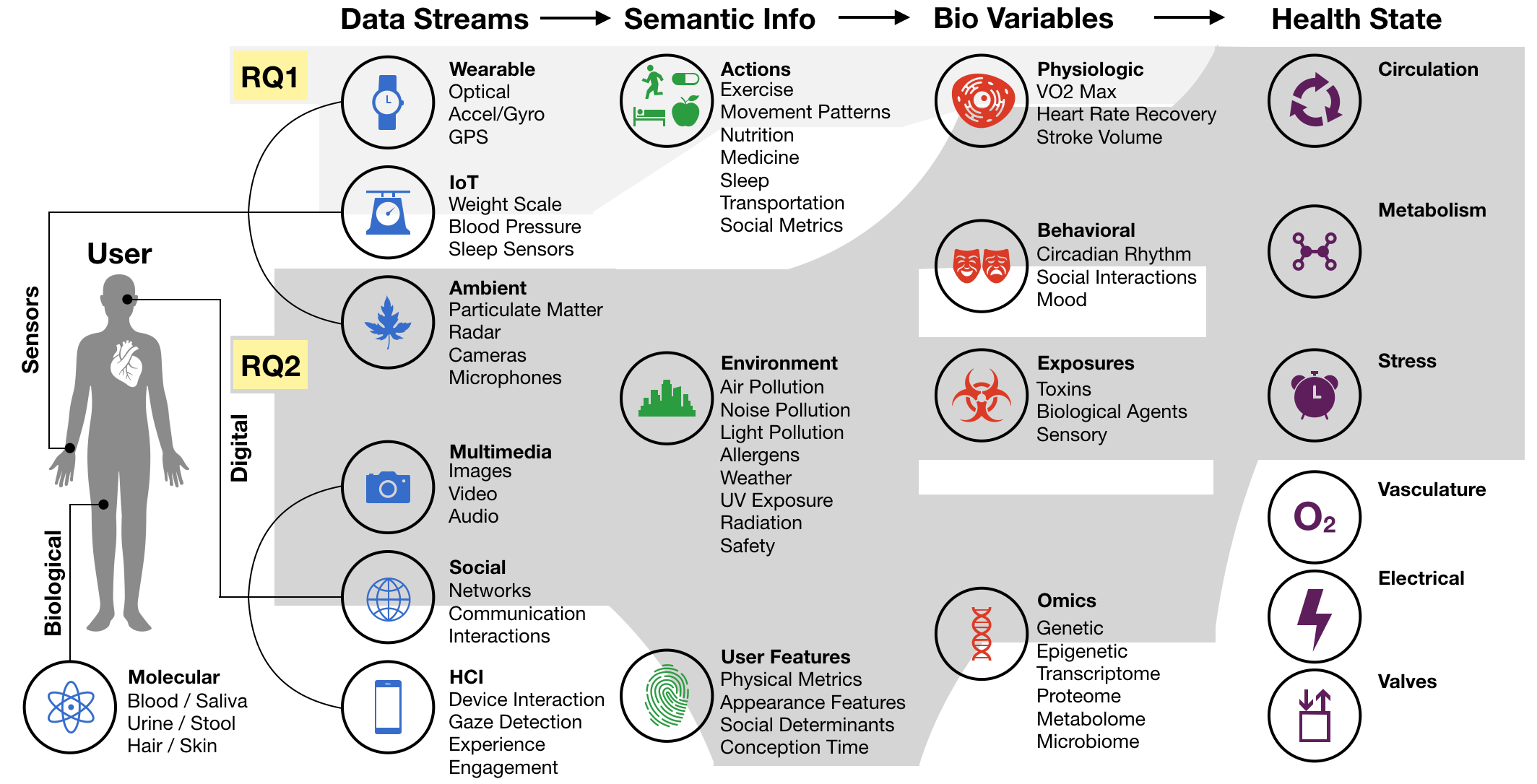}
\caption{Intermediate steps of transforming cross-modal data into a bio-variables or health state (for cardiovascular health in this work). Research question 1 delves into taking wearable data streams and producing a single biological variable of VO2 Max (CRF equivalent). Research question 2 takes a much larger set of data streams to produce multiple biological variables that can then be used to approximate a health state of the individual. Unshaded additional data streams, information, biological variables, and health states are not addressed in detail but shown to demonstrate future potential.}~\label{fig:flow}
\vspace{-3mm}
\end{figure*}

Transforming this data into semantically meaningful information is the first step in using data for an end goal. In the application of health, an additional step is needed to take these information bits into the domain of biological variables. A sufficient set of biological variables can provide an overview of how the human system is operating. This flow is shown in Figure \ref{fig:flow}.

\subsection{RQ1: Estimation of CRF Bio-Variable}

CRF represents the integrated biological performance of delivering oxygen from the atmosphere via the lungs and blood to the mitochondria to perform physical work (Work = Force x Distance). This essentially quantifies the functional capacity of the respiratory, cardiovascular, metabolic, and type-1 fiber musculature. CRF is usually measured through breath captured maximal oxygen consumption (VO2max) during a maximal exercise effort of several minutes. Because the efficiency of muscular work produced per unit oxygen consumed is directly related to the physics of adenosine triphosphate synthesis and breakdown to adenosine diphosphate, we can use power output on bicycle ergometry to directly calculate VO2max \cite{Maier2017AccuracyCycling,Lucia2002KineticsCyclists}. VO2max is measured in mL oxygen consumed / minute / kilogram of bodyweight and is a direct measurement of CRF. For the purposes of this paper, they are equivalent.

\begin{figure}
\small
\centering
\includegraphics[width=.95\columnwidth]{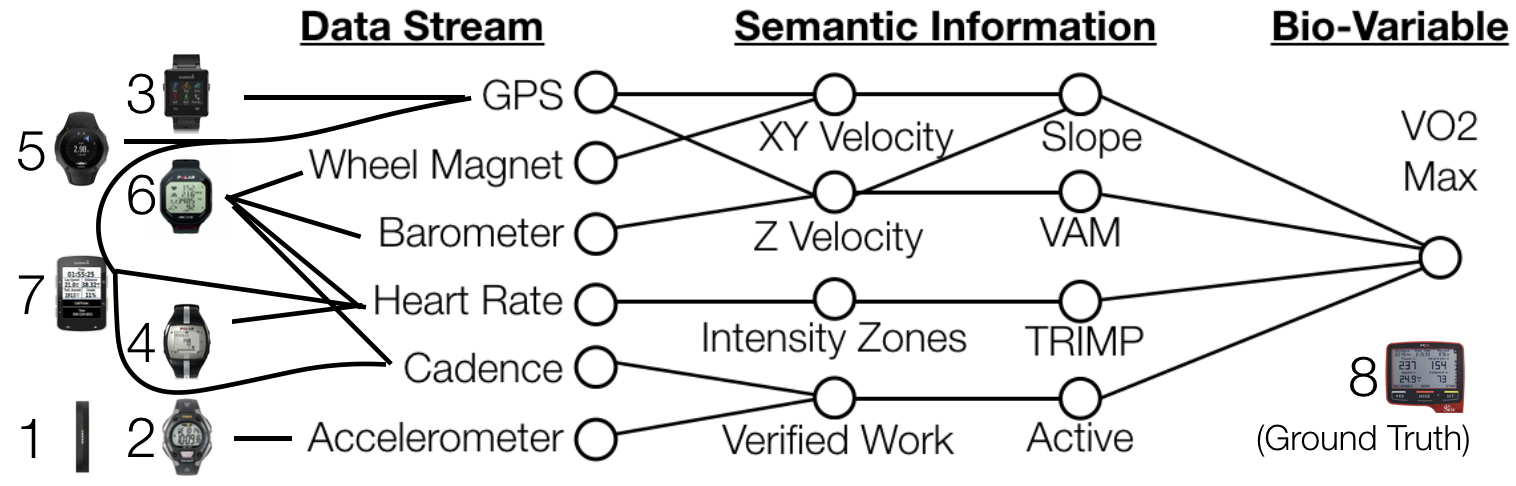}
\caption{Wearable devices used for comparison in the experiments and respective feature extraction: 1. Timex Ironman, 2. Fitbit Flex2, 3. Garmin VivoActive, 4. Polar FT7, 5. Suunto Spartan, 6. Polar RCX5, 7. Garmin Edge 520, 8. SRM-PC8 (contains all sensors and used for ground truth).}~\label{fig:device}
\vspace{-5mm}
\end{figure}

We use a multi step process to extract meaningful information from the wearable devices (Figure (\ref{fig:device}). Forces against a bicycle motion in real world activity are divided into three main components: wind, gravity, and friction. Effort by a rider can be measured by duration of exercise or heart rate based effort. We use three methods to estimate power output, and test these methods against the ground truth of known power output from device 8.
\newline
\textbf{Active Time Based Training Effect - TIME:} We use devices 1 and 2 and instances where we only have accelerometer, time, or cadence data to estimate how much time the user is actively exercising. We base this estimate from the increased exercise volume (time) leading to increased CRF \cite{Jones2000TheFitness}.
\newline
\textbf{Heart Rate Based Training Effect - TRIMP:} Devices 4,5,6, and 7 have heart rate sensors we use to predict not only exercise volume, but also intensity. Intensity of exercise is calculated by the established Training Impulse (TRIMP) method \cite{Banister1980PlanningTraining.}.
\newline
\textbf{Work Against Gravity - VAM:} We use devices 3,5,6,and 7 to give us both horizontal and vertical velocity. Devices 3,5 and 7 use GPS to give latitude,longitude and altitude. Device 6 uses a barometer for altitude and wheel magnet for horizontal velocity. Vertically Ascended Meters (VAM) is the z-axis velocity in meters/hour. For all instances where the rider is going uphill, we calculate the Newtonian physical work done against gravity. The horizontal velocity is less climbing uphill, and thus we assume a minimal component of wind resistance.

We test these estimation methods to compare performance in prediction of CRF with different situations of sensor derived information. First, we use a global prediction model by using a subset of 50\% of subjects. We use an individual model for instances where a user would be given a calibration device of a power meter for a given period of time, and measure how well we can model future CRF prediction after the calibration device was removed. For both instances we use 70\% of the data for training.

\begin{figure}
\small
\centering
\includegraphics[width=1\columnwidth]{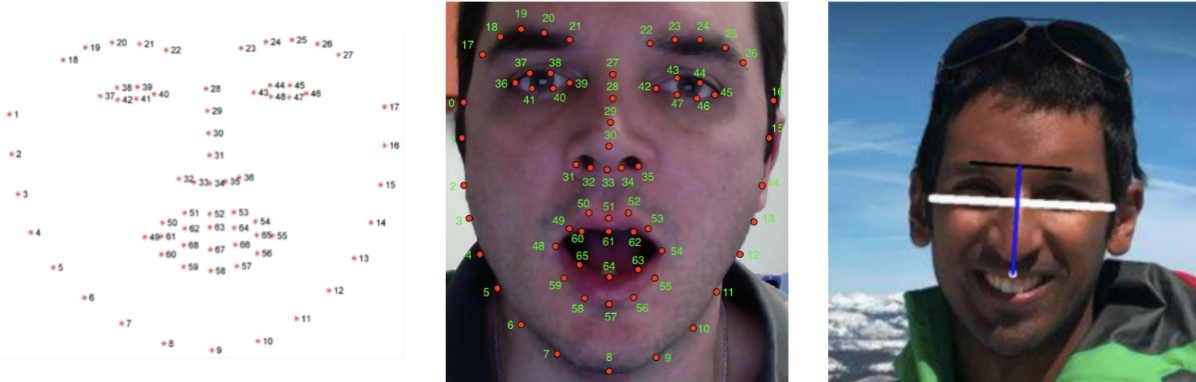}
\caption{Images of people provide insight into their health state. Visual features of facial width-height ratio used as a validated proxy for basal genetic testosterone levels.}~\label{fig:face}
\vspace{-5mm}
\end{figure}

\begin{figure}
\small
\centering
\includegraphics[width=1.02\columnwidth]{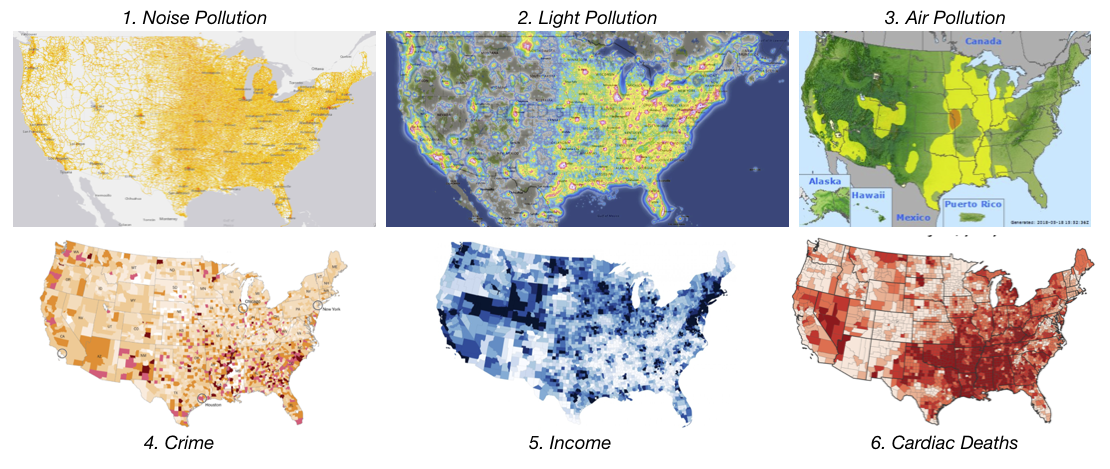}
\caption{Visualization of various geo-spatial data sources used in our health state estimation.}~\label{fig:maps}
\vspace{-5mm}
\end{figure}

\subsection{RQ2: Multi-Factorial Approach to Health}

The true state of the cardiovascular system will depend upon many different controlled inputs and disturbances. In this section we provide an example of how varied these sources can be, and how they may be integrated to provide a dashboard of the cardiovascular state to a user or health expert. The total state of cardiovascular health may be summarized into some sub-states such as circulation, metabolism, stress, vascular perfusion, electrical activity, and valvular function for example. As shown in Figure \ref{fig:flow}, these sub-states are composed from evidence based relationships with bio-variables. We describe the methods used to gather and extrapolate these data relationships below:
\newline
\textbf{Images} of the user were used to derive several biological features. We use OpenCV facial landmarks detection to determine width-to-height ratio as shown in Figure \ref{fig:face}  \cite{OpenCVFacialDetection}. We calculate a proxy for genetic testosterone levels through this ratio \cite{Lefevre2013TellingMen}. Higher testosterone levels are positively correlated with better cardiovascular circulation and metabolism \cite{Oskui2013TestosteroneLiterature}. We also use the images for ethnicity detection and gender identification.
\newline
\textbf{Location} of living for each user was determined through histograms of GPS coordinates at the beginning of the activities. These locations were mapped to zip codes in the United States.
\newline
\textbf{Environmental} Zip code and county based average income, cardiac deaths, community crime risk \cite{UnitedStatesofAmericaCentersPrevention}, air \cite{UnitedStatesofAmericaAirNowAgency}, light\cite{UnitedStatesofAmerica-NOAALightMap},  and noise pollution \cite{UnitedStatesofAmerica-DepartofTransportationNationalStatistics} data were then mapped to each user. Established relationships that affect cardiovascular health have been reported for PM2.5 air pollution \cite{Sun2010CardiovascularExposure}, light pollution and noise pollution \cite{Munzel2018EnvironmentalSystem}.
\newline
\textbf{Circadian} light exposure during exercise were derived from physics based models of earth rotation to determine natural light exposure. Patterns of weekly variability in exercise habits and time zone changes were also calculated from user data, and light pollution at the living location. Circadian disruption has been shown in humans to cause cardiovascular impairment \cite{Scheer2009AdverseMisalignment}.
\newline
\textbf{Social Media} networks were used to not only gather the images, but also professional status and educational attainment of the user from LinkedIn. The combination of education, professional status, and zip code average income was used to estimate financial status \cite{AugustB.Hollingshead1975FourStatus}. Education links to cardiovascular disease as a proxy for other risk factors have been studied \cite{Degano2017TheIndex,Kubota2017AssociationStudy}.
\newline
\textbf{Surveys} were given to users to obtain their age, smoking status, height, weight, and waist circumference. These measurements can also be automated with IoT devices, such as connected weight scales. Body Mass Index (BMI) \cite{Wilson2002OverweightRisk} and Waist-to-Height Ratio (WHR) \cite{Ashwell2016Waist-to-heightCircumference} were derived from these values.
\newline
\textbf{Wearables} (specifically device 8) from RQ1 were used to estimate bio-variables that have relationships with cardiovascular disease and heart functionality. These bio-variables include heart rate recovery \cite{Cole1999Heart-RateMortality}, heart left ventricle stroke volume \cite{Astrand1964CardiacWork.}, heart rate drift\cite{Coyle2001CardiovascularPerspectives}, kilojoules of work \cite{Hamilton2007RoleDisease} in addition to the CRF, TRIMP, and active time.

Inherent ASCVD risk was calculated from a combination of established risk factors due to ethnicity, age, basal testosterone, and smoking status. This is an established risk of potential for a hard cardiovascular event \cite{Goff20142013Guidelines} within the next 10 years for that individual.
We define high friction risk as factors that require dramatic life change to alter (such as moving to a new home or acquiring a higher educational degree). In our case this relates to the variables of education attainment and income in addition to factors related to living location which include crime, local incidence of cardiac death, air, noise, and light pollution. Circadian Rhythm Disruption is a normalized sum of light pollution in living location, time zone changes (hours changed relative to GMT in last 4 weeks), and exercise habit variability (average exercise start time difference from previous day in last 4 weeks). Circulation capability of the heart is a normalized sum of CRF, HRR, SV, TRIMP in the last 4 weeks. These factors capture the ability of the heart to pump blood throughout the body. Metabolism summary was calculated as the normalized sum of exercise work (in kJ), HRD, active time, BMI and WHR. These factors capture the ability of the individual to maintain high resting basal metabolic rate and resist fatigue. These summaries does not reflect any absolute risk, just relative risk to others in our sample subjects. They are meant to observe how an individual's health status is longitudinally changing over time, or as a cross-sectional comparison to others in the same subject population. The overall heart score is an equal weighted average of both these relative metrics and the inherent ASVCD risk.

Multi-modal data assimilation and visualizations have been used extensively to maintain the health state of jet engines and other mechanical devices \cite{Simon2004SensorEngines}. By placing various sensors on the engine, engineers and pilots are able to monitor the status of an engine in real-time and understand when to take precaution or perform an action to ensure the safety and longevity of the engine. We present a similar view of health data in Figure \ref{fig:mobile} for individual use and Figure \ref{fig:multi} for professional/expert use.

\section{Experimental Results}

The dataset used in the experiments includes sensor data streams at one second resolution from eight wearable devices collected on 24 male cycling athletes over an average of 5 years in the United States. Athletes also had strain gauges installed on their bicycles to measure physiologic true power output. The total dataset includes 31,776 activities and 70,178 hours of exercise data. Social media outlets of Instagram and LinkedIn were also used to gather an image dataset of 50 images per athlete and general demographic background information. Environmental data was sourced from government or open source databases.

\subsection{RQ1: CRF Bio-Variable Estimation}
\begin{figure}
\small
\centering\includegraphics[width=1 \columnwidth]{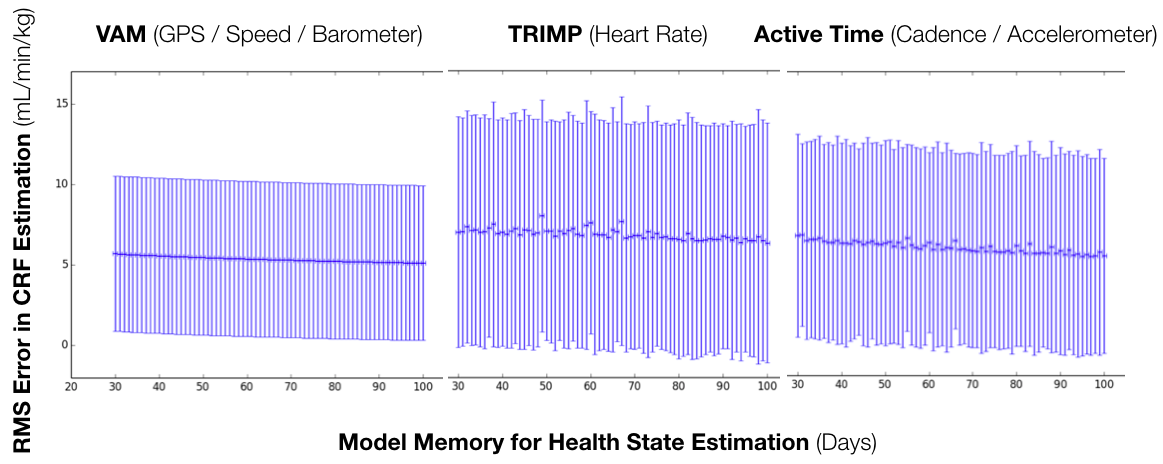}
\caption{Comparison of sensor prediction performance based on changing the memory in the model in determining health state. Based on a p-value of 0.05, there were no statistically significant differences in choosing the memory value. Thus, we chose the established standard of 42 days for our model \cite{Jones2000TheFitness}. }~\label{fig:parameter}
\vspace{-7mm}
\end{figure}

\begin{figure}
\small
\centering\includegraphics[width=1 \columnwidth]{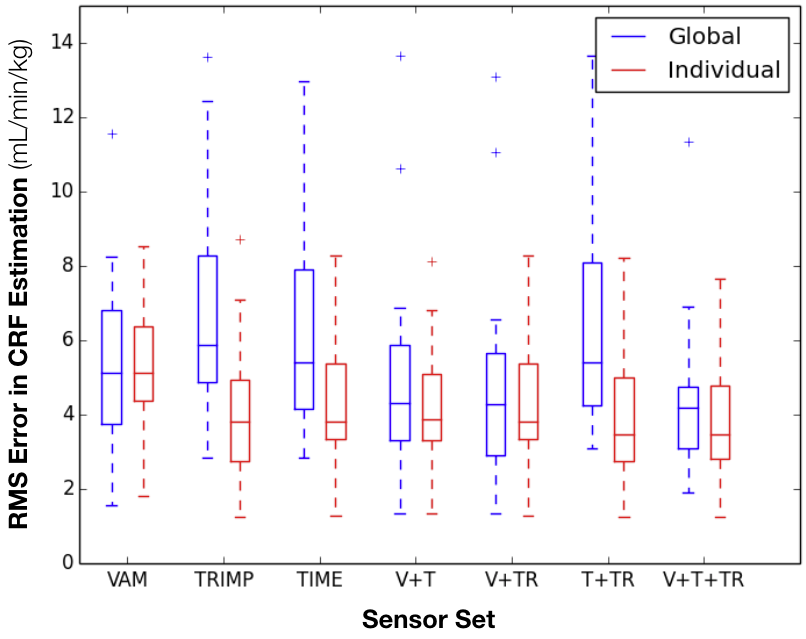}
\caption{Comparison of sensor prediction performance. Global models are trained on all data collected from a subset of subjects. Individual models are trained on data collected from one subject. Global models can be used for estimation before the individual model for a person is calibrated. }~\label{fig:box}
\vspace{-7mm}
\end{figure}

% In the first set of experiments, our goal is to estimate a bio-variable using data captured from multimodal sensors, and show that the estimations can be improved by combining multiple cheap (in terms of monetary or energy costs) sensors.
% We have used 3 features: 1) Rate of vertical ascent (VAM), 2) Training Impulse (TRIMP) and 3) Active Time, which have already been established to have a direct impact on CRF. 
% \newline
Per second power output values collected from the strain gauges were used as ground truth in our experiments for bio-variable estimation. We used a rolling average of maximum 4 minute power output per day over 42 days to generate the ground truth for our experiments.
We trained two sets of linear regression models for each feature, a global model and a personal model. The global model was trained using the data collected from a subset of subjects and tested on the remaining subjects. The personal model for each individual was trained on a 70\% training subset for the subject and tested on the remaining 30\% subset.
\newline
\textbf{VAM} models are trained to predict average power output (normalized by body weight) in 4 minute windows in an activity using the VAM in the time window, and the maximum estimated power output is then used to compute a continuous daily estimate for VO2 Max. We trained models with varying slope thresholds to identify the impact of slope on estimate accuracy. As the slope increases, the effect of other resistance factors (such as wind, rolling resistance) decreases and the model performs better (Table \ref{table:slope_error}). We choose which model to use based on the maximum slope observed in the 4 minute windows, for example if in a ride the maximum slope observed in a 4 minute interval is 5.3\%, we would choose the model trained on intervals where slope is greater than 5\%. Since we are predicting body weight normalized power using VAM, none of the two metrics are greatly influenced by individual parameters. This is reflected in similar global and individual model performances for VAM (fig \ref{fig:box}).
\newline
\textbf{TRIMP} captures the work done by an individual's heart in the last 42 days. We trained linear regression model to predict an individual's VO2 Max value based on their total TRIMP score in past 42 days. This metric proved to be more effective in a personal model than a global model as different individuals have different heart rate response to same exercise intensity (fig \ref{fig:box}).
\newline
\textbf{Active time} is the actual amount of time the individual was actively putting in effort in past 42 days. We obtained this metric using cadence values collected at per second resolution. We trained linear regression model to predict an individual's VO2 Max value based on their total activity time in past 42 days. Similar to TRIMP, this metric performs better in a personalized model than in global model as different individuals would have a different response to the same exercise volume (fig. \ref{fig:box}).
\newline
\textbf{Combination models} have outperformed their constituent models in all our experiments as shown by the error plots in fig. \ref{fig:box}. The estimates from the previous models were combined using a weighted average, where weights for a model estimate are inverse of the model's training error. The error in estimates for these models are reported in fig. \ref{fig:box} and discussed in this section.
We can see from the plot that the best model in terms of average error and variance in error utilizes all available data streams.
\newline
We also performed an experiment to find out the optimum time to be considered for aggregating the metrics while estimating CRF values. We plotted the test error for the global models utilizing one metric in fig. \ref{fig:parameter}. We can see that while there is some variation in mean error, the 95\% confidence intervals overlap for all time windows and we cannot find an optimum time window to use in our experiments based solely on the data. Therefore we have used the clinically recommended period of 42 days to aggregate the past exercise events.

\begin{table}
\centering
\caption{Slope based optimization of VAM models}~\label{table:slope_error}
\resizebox{\columnwidth}{!}{%
\begin{tabular}{|l|l|l|l|l|} 
\hline
\begin{tabular}[c]{@{}l@{}}Slope~\\threshold (\%)\end{tabular} & \begin{tabular}[c]{@{}l@{}}Test Set RMSE~\\(Rel. Power)\end{tabular} & \begin{tabular}[c]{@{}l@{}}Training Set~RMSE~\\(Rel. Power)\end{tabular} & \begin{tabular}[c]{@{}l@{}}Training~\\R Squared\end{tabular} & \begin{tabular}[c]{@{}l@{}}Size of~\\training set \end{tabular}  \\ 
\hline
0+                                                             & 0.726                                                                & 0.665                                                                    & 0.381                                                        & 16810792                                                         \\
1+                                                             & 0.620                                                                & 0.537                                                                    & 0.527                                                        & 10728610                                                         \\
2+                                                             & 0.557                                                                & 0.473                                                                    & 0.593                                                        & 7952334                                                          \\
3+                                                             & 0.488                                                                & 0.424                                                                    & 0.655                                                        & 6236507                                                          \\
4+                                                             & 0.451                                                                & 0.391                                                                    & 0.695                                                        & 4884243                                                          \\
5+                                                             & 0.420                                                                & 0.363                                                                    & 0.732                                                        & 3509724                                                          \\
6+                                                             & 0.405                                                                & 0.344                                                                    & 0.760                                                        & 2294199                                                          \\
7+                                                             & 0.395                                                                & 0.328                                                                    & 0.781                                                        & 1400488                                                          \\
8+                                                             & 0.365                                                                & 0.318                                                                    & 0.793                                                        & 781841                                                           \\
9+                                                             & 0.347                                                                & 0.317                                                                    & 0.797                                                        & 423315                                                           \\
\hline
\end{tabular}
}
\vspace{-4mm}
\end{table}

\subsection{RQ2: Cross-Modal Heart Health State}
Even while referencing established bioscience research we still have no way to validate ground truth until decades into the future when people die. So comparisons are not the best way to experimentally validate this question. Performance comparisons for this type of experiment will need to be validated through large scale data collection and monitoring in prospective studies as mentioned in the related works. This research question largely poses a beginning for how multi-factorial health states can begin to surface for use.

\begin{figure*}
\small
\centering
 \includegraphics[width=1.9 \columnwidth]{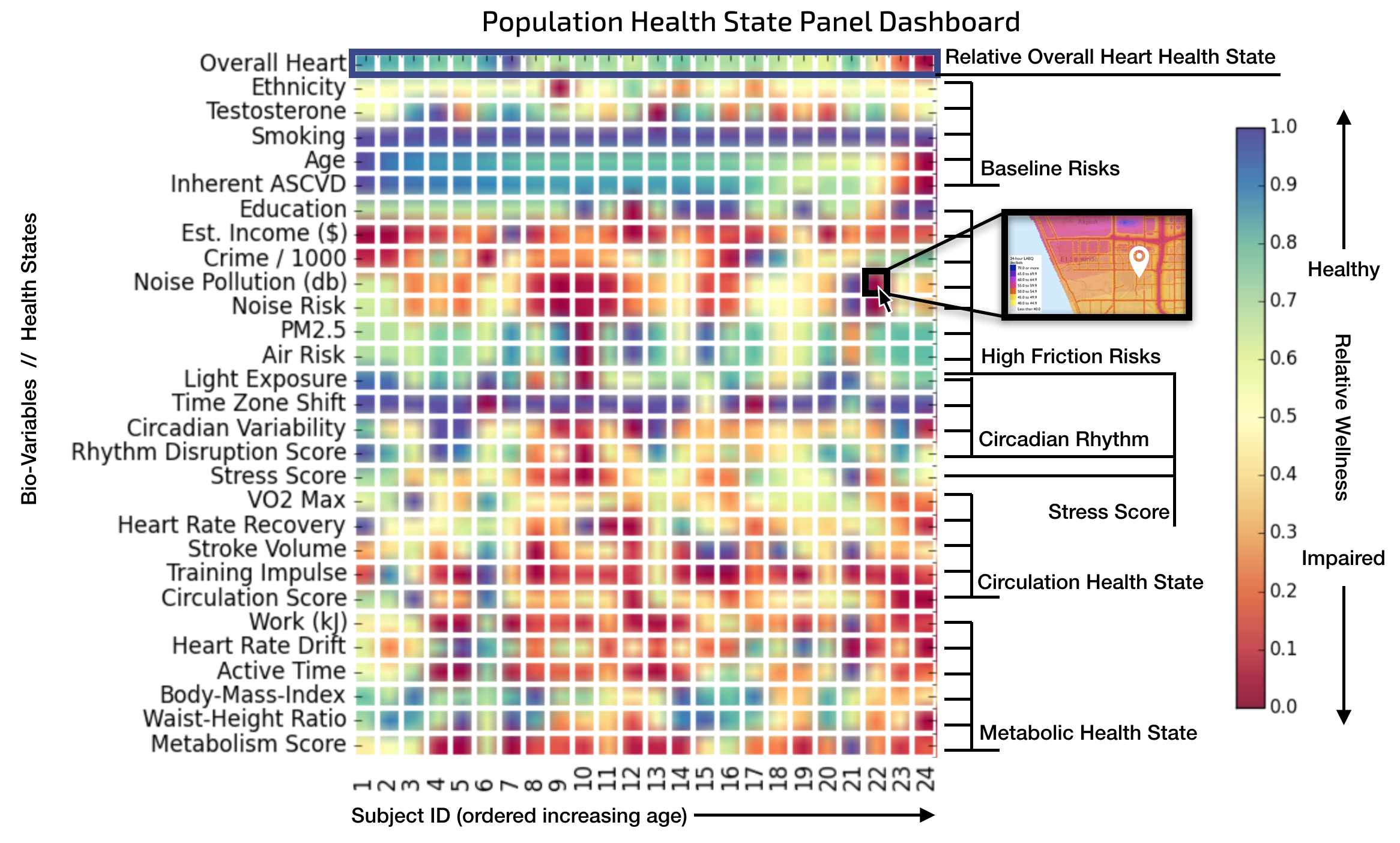}
\caption{Heat map of bio-variables and summary scores that affect each individual subject. This type of visualization integrates cross-modal data in a manner that a clinician, hospital, public health agency, or any expert can use to monitor health of a patient panel. Clicking on a certain box would pull up further insights and details.}~\label{fig:multi}
\vspace{-6mm}
\end{figure*}

Using the approach described in section 4, we assimilate various biological parameters for each of the 24 subjects as shown in Figure \ref{fig:multi}. We find that even though most of these subjects are all cycling athletes, they have a wide range in both their bio-variables and environmental exposures. Current day primary care doctors would not be able to see this when a patient visits.

\begin{figure}
\small
\centering
\includegraphics[width=.95 \columnwidth]{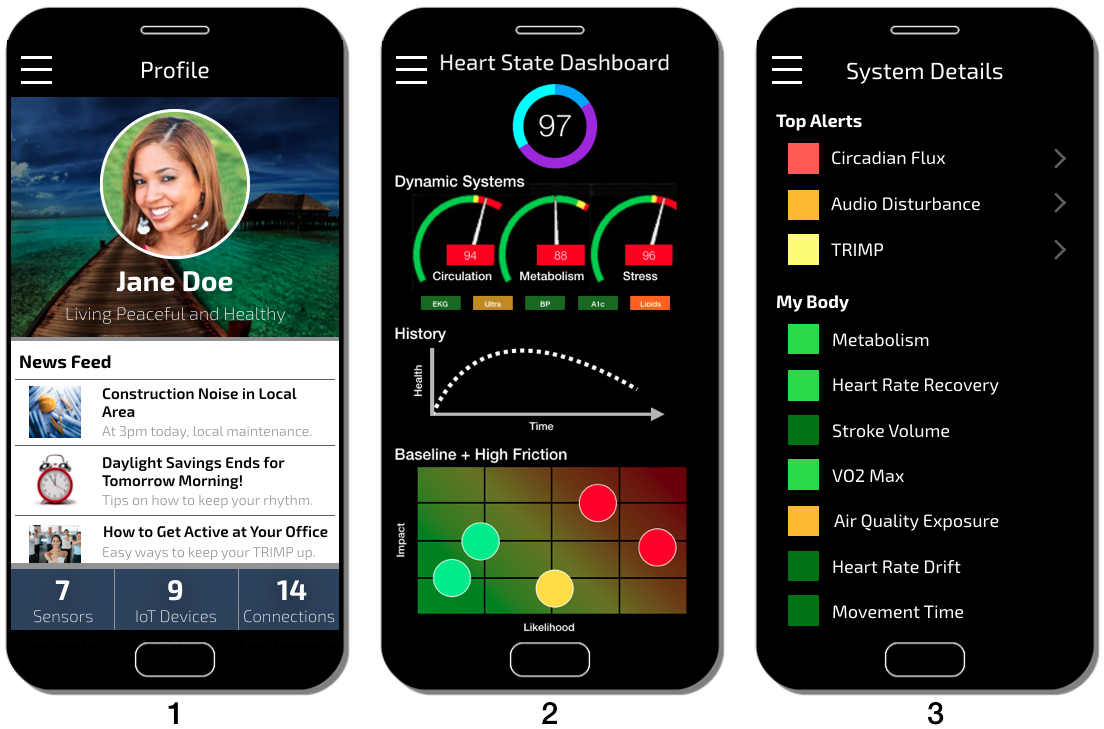}
\caption{Screen 1 we show how health state can fuel recommendations. Inspired by jet engine dashboards, screen 2 can give a live snapshot of the health state. Screen 3 gives a comprehensive list of all bio-variables and health states being tracked, with a ranking system to provide relevant results at the top.}~\label{fig:mobile}
\vspace{-5mm}
\end{figure}

This data assimilation can be used inform personal health state as shown in Figure \ref{fig:mobile}. The combination of sensors, IoT devices, and environmental data connections can provide a rich experience to interact with meaningful health insights. This can be for individual use, or for use in when a user visits a health care provider. One step further would be to link this with their electronic medical record system. Looking at this data panel across the subject pool, we discover some interesting trends in Figure \ref{fig:multi}. As expected, as the age increases the overall heart health state decreases, since age is a large factor of cardiac health (age range in panel is 18-57). As age increases, we also see a reduction in crime and noise pollution, suggesting that older individuals live in safer and quieter neighborhoods. We also see circadian rhythm disruption maximally in the middle ages (20-29), suggested a more erratic lifestyle for those in their twenties. Circulation and metabolism scores also trend (including VO2 Max / CRF) lower as age increases. Although this is a small sample size to make any strong conclusions, we can begin to see the power of using this cross data analysis for health.

\section{Conclusions}
In this paper we propose an approach to leverage cross-modal data to estimate a needed health variable or health state in the context of cybernetics. Specifically in the focus of cardiovascular health, we estimate CRF from various wearable devices. From our experimental results we can see that increasing the number of data streams provide increased performance characteristics in achieving this goal. Furthermore, we show that total health of an individual is much greater than any single biological variable, and that we need to integrate a diverse array of data types to more better understand the total health state of a particular individual or organ system. Ideally we have some actionable or semantically meaningful dashboard as shown in Figures \ref{fig:mobile} and \ref{fig:multi}, which a user or a health expert can reference to get an "engine check" of the health state in real time.
\newline
\textbf{Implications:}
The proposed utility of this work is to open the concept of using a diverse array of data streams to improve health state estimation. In the ideal case, this also lowers the cost for instantaneous health assessment, and provides increased value for individuals to purchase sensors like wearables and IoT devices. Individuals may be more motivated to track their health state, especially if it will be used in professional clinical decision making or influencing daily actions. This also provides the user with instant feedback with results from their lifestyle modification, medicine, environment and more. Perhaps this may be used as a tool to encourage healthy habits, or to avoid dangerous environments.
\newline
\textbf{Limitations:}
Estimation in its initial iteration may not be accurate, but it is assumed to improve over time with refining of equations, algorithms, feature extraction methods, learning methods, as well as with improvements in hardware technology. We use linear methods in this paper as a starting point for obvious further advancement with advanced learning and predictive methods. Baseline comparisons are also difficult when studying individual subjects, and will require statistical methods for n of 1 studies \cite{Senn2017SampleTrials,Sedgwick2014WhatTrial,Gabler2011N-of-1Review,Duan2013Single-patientResearch}. Wearable devices and similar low cost sensors are currently better used as a screening tool to identify if a user is at risk or their health state is changing, but clinical gold standard testing (which is more expensive) might be used to confirm the true state of the user if the situation is critical. More work will need to be done to ensure the robustness of the estimations are good enough to use alone for important clinical decisions from more validated clinical and biomedical research. This work at the current stage cannot give a validated prediction window of when an adverse event (eg. myocardial infarction) may happen. 
\newline
\textbf{Future Directions:}
We hope to show how wearable devices, Iot, images, along with other data types can potentially be used in lieu of expensive sensors to estimate health status. For a single device, we show how multiple types of sensors improves prediction quality. Beyond a single device, we show that an assimilation of more diverse data with domain knowledge can further illustrate a wide view of health states. Additionally, privacy and security methods must evolve concurrently for such systems to function in the real world. Estimation is just one part of the cybernetic health paradigm. Ultimately, work must be done to ensure that reliable and useful systems are developed to guide people towards better health. Our dataset will be available in the public domain to encourage other researchers. We invite others to participate in building the foundational blocks of health state estimation and cybernetic health so that we can all enjoy a more informed and healthier life.

\bibliographystyle{ACM-Reference-Format}
\balance
\bibliography{megi}


\begin{thebibliography}{00}

%%% the Web address.

\ifx \showCODEN    \undefined \def \showCODEN     #1{\unskip}     \fi
\ifx \showDOI      \undefined \def \showDOI       #1{#1}\fi
\ifx \showISBNx    \undefined \def \showISBNx     #1{\unskip}     \fi
\ifx \showISBNxiii \undefined \def \showISBNxiii  #1{\unskip}     \fi
\ifx \showISSN     \undefined \def \showISSN      #1{\unskip}     \fi
\ifx \showLCCN     \undefined \def \showLCCN      #1{\unskip}     \fi
\ifx \shownote     \undefined \def \shownote      #1{#1}          \fi
\ifx \showarticletitle \undefined \def \showarticletitle #1{#1}   \fi
\ifx \showURL      \undefined \def \showURL       {\relax}        \fi
% invisible to TeX
\providecommand\bibfield[2]{#2}
\providecommand\bibinfo[2]{#2}
\providecommand\natexlab[1]{#1}
\providecommand\showeprint[2][]{arXiv:#2}

\bibitem[\protect\citeauthoryear{Albinali, Intille, Haskell, and
  Rosenberger}{Albinali et~al\mbox{.}}{2010}]%
        {Albinali2010UsingEstimation}
\bibfield{author}{\bibinfo{person}{Fahd Albinali}, \bibinfo{person}{Stephen
  Intille}, \bibinfo{person}{William Haskell}, {and} \bibinfo{person}{Mary
  Rosenberger}.} \bibinfo{year}{2010}\natexlab{}.
\newblock \showarticletitle{{Using wearable activity type detection to improve
  physical activity energy expenditure estimation}}.
\newblock \bibinfo{journal}{{\em Proceedings of the 12th ACM international
  conference on Ubiquitous computing - Ubicomp '10\/}} (\bibinfo{year}{2010}),
  \bibinfo{pages}{311}.
\newblock
\showISBNx{9781605588438}
\showDOI{%
\url{https://doi.org/10.1145/1864349.1864396}}


\bibitem[\protect\citeauthoryear{Altini, Casale, Penders, and Amft}{Altini
  et~al\mbox{.}}{2015}]%
        {Altini2015PersonalizedModels}
\bibfield{author}{\bibinfo{person}{Marco Altini}, \bibinfo{person}{Pierluigi
  Casale}, \bibinfo{person}{Julien Penders}, {and} \bibinfo{person}{Oliver
  Amft}.} \bibinfo{year}{2015}\natexlab{}.
\newblock \showarticletitle{{Personalized cardiorespiratory fitness and energy
  expenditure estimation using hierarchical Bayesian models}}.
\newblock \bibinfo{journal}{{\em Journal of Biomedical Informatics\/}}
  \bibinfo{volume}{56} (\bibinfo{year}{2015}), \bibinfo{pages}{195--204}.
\newblock
\showISBNx{1532-0480 (Electronic) 1532-0464 (Linking)}
\showISSN{15320464}
\showDOI{%
\url{https://doi.org/10.1016/j.jbi.2015.06.008}}


\bibitem[\protect\citeauthoryear{Altini, Casale, Penders, and Amft}{Altini
  et~al\mbox{.}}{2016}]%
        {Altini2016CardiorespiratorySensors}
\bibfield{author}{\bibinfo{person}{Marco Altini}, \bibinfo{person}{Pierluigi
  Casale}, \bibinfo{person}{Julien Penders}, {and} \bibinfo{person}{Oliver
  Amft}.} \bibinfo{year}{2016}\natexlab{}.
\newblock \showarticletitle{{Cardiorespiratory fitness estimation in
  free-living using wearable sensors}}.
\newblock \bibinfo{journal}{{\em Artificial Intelligence in Medicine\/}}
  \bibinfo{volume}{68} (\bibinfo{year}{2016}), \bibinfo{pages}{37--46}.
\newblock
\showISBNx{09333657 (ISSN)}
\showISSN{18732860}
\showDOI{%
\url{https://doi.org/10.1016/j.artmed.2016.02.002}}


\bibitem[\protect\citeauthoryear{Ashwell and Gibson}{Ashwell and
  Gibson}{2016}]%
        {Ashwell2016Waist-to-heightCircumference}
\bibfield{author}{\bibinfo{person}{Margaret Ashwell} {and}
  \bibinfo{person}{Sigrid Gibson}.} \bibinfo{year}{2016}\natexlab{}.
\newblock \showarticletitle{{Waist-to-height ratio as an indicator of early
  health risk: Simpler and more predictive than using a matrix based on BMI and
  waist circumference}}.
\newblock \bibinfo{journal}{{\em BMJ Open\/}} \bibinfo{volume}{6},
  \bibinfo{number}{3} (\bibinfo{year}{2016}).
\newblock
\showISSN{20446055}
\showDOI{%
\url{https://doi.org/10.1136/bmjopen-2015-010159}}


\bibitem[\protect\citeauthoryear{Astrand, Cuddy, Saltin, and Stenberg}{Astrand
  et~al\mbox{.}}{1964}]%
        {Astrand1964CardiacWork.}
\bibfield{author}{\bibinfo{person}{Per-Olof Astrand}, \bibinfo{person}{T~Edward
  Cuddy}, \bibinfo{person}{Bengt Saltin}, {and} \bibinfo{person}{Jesper
  Stenberg}.} \bibinfo{year}{1964}\natexlab{}.
\newblock \showarticletitle{{Cardiac Output During Submaximal and Maximal
  Work.}}
\newblock \bibinfo{journal}{{\em J Appl Physiol\/}}  \bibinfo{volume}{19}
  (\bibinfo{year}{1964}), \bibinfo{pages}{268--274}.
\newblock
\showISBNx{8750-7587}
\showISSN{0021-8987}


\bibitem[\protect\citeauthoryear{{August B. Hollingshead}}{{August B.
  Hollingshead}}{1975}]%
        {AugustB.Hollingshead1975FourStatus}
\bibfield{author}{\bibinfo{person}{{August B. Hollingshead}}.}
  \bibinfo{year}{1975}\natexlab{}.
\newblock \showarticletitle{{Four Factor Index of Social Status}}.
\newblock \bibinfo{journal}{{\em YALE JOURNAL OF SOCIOLOGY\/}}
  \bibinfo{volume}{8} (\bibinfo{year}{1975}).
\newblock
\showURL{%
\url{https://s3.amazonaws.com/academia.edu.documents/30754699/yjs_fall_2011.pdf?AWSAccessKeyId=AKIAIWOWYYGZ2Y53UL3A&Expires=1526782500&Signature=zI74sfIS0kLlqQq4aYRtpnLp2NI%3D&response-content-disposition=inline%3B%20filename%3DAugust_B._Hollingshead_s_Four_Fa}}


\bibitem[\protect\citeauthoryear{Banister and Calvert}{Banister and
  Calvert}{1980}]%
        {Banister1980PlanningTraining.}
\bibfield{author}{\bibinfo{person}{E~W Banister} {and} \bibinfo{person}{T~W
  Calvert}.} \bibinfo{year}{1980}\natexlab{}.
\newblock \showarticletitle{{Planning for future performance: implications for
  long term training.}}
\newblock \bibinfo{journal}{{\em Canadian journal of applied sport sciences.
  Journal canadien des sciences appliquees au sport\/}} \bibinfo{volume}{5},
  \bibinfo{number}{3} (\bibinfo{date}{9} \bibinfo{year}{1980}),
  \bibinfo{pages}{170--6}.
\newblock
\showISSN{0700-3978}
\showURL{%
\url{http://www.ncbi.nlm.nih.gov/pubmed/6778623}}


\bibitem[\protect\citeauthoryear{Borresen and Ian~Lambert}{Borresen and
  Ian~Lambert}{2009}]%
        {Borresen2009ThePerformance}
\bibfield{author}{\bibinfo{person}{Jill Borresen} {and}
  \bibinfo{person}{Michael Ian~Lambert}.} \bibinfo{year}{2009}\natexlab{}.
\newblock \showarticletitle{{The Quantification of Training Load, the Training
  Response and the Effect on Performance}}.
\newblock \bibinfo{journal}{{\em Sports Medicine\/}} \bibinfo{volume}{39},
  \bibinfo{number}{9} (\bibinfo{date}{9} \bibinfo{year}{2009}),
  \bibinfo{pages}{779--795}.
\newblock
\showISSN{0112-1642}
\showDOI{%
\url{https://doi.org/10.2165/11317780-000000000-00000}}


\bibitem[\protect\citeauthoryear{Cole, Blackstone, Pashkow, Snader, and
  Lauer}{Cole et~al\mbox{.}}{1999}]%
        {Cole1999Heart-RateMortality}
\bibfield{author}{\bibinfo{person}{Christopher~R. Cole},
  \bibinfo{person}{Eugene~H. Blackstone}, \bibinfo{person}{Fredric~J. Pashkow},
  \bibinfo{person}{Claire~E. Snader}, {and} \bibinfo{person}{Michael~S.
  Lauer}.} \bibinfo{year}{1999}\natexlab{}.
\newblock \showarticletitle{{Heart-Rate Recovery Immediately after Exercise as
  a Predictor of Mortality}}.
\newblock \bibinfo{journal}{{\em New England Journal of Medicine\/}}
  \bibinfo{volume}{341}, \bibinfo{number}{18} (\bibinfo{year}{1999}),
  \bibinfo{pages}{1351--1357}.
\newblock
\showISBNx{0028-4793 (Print){\textbackslash}r0028-4793 (Linking)}
\showISSN{0028-4793}
\showDOI{%
\url{https://doi.org/10.1056/NEJM199910283411804}}


\bibitem[\protect\citeauthoryear{Coyle and Gonz{\'{a}}lez-Alonso}{Coyle and
  Gonz{\'{a}}lez-Alonso}{2001}]%
        {Coyle2001CardiovascularPerspectives}
\bibfield{author}{\bibinfo{person}{E~F Coyle} {and} \bibinfo{person}{J.
  Gonz{\'{a}}lez-Alonso}.} \bibinfo{year}{2001}\natexlab{}.
\newblock \showarticletitle{{Cardiovascular Drift during Prolonged Exercise:
  New Perspectives}}.
\newblock \bibinfo{journal}{{\em Exercise and Sport Sciences Reviews\/}}
  \bibinfo{volume}{29}, \bibinfo{number}{2} (\bibinfo{year}{2001}),
  \bibinfo{pages}{88--92}.
\newblock
\showISBNx{0091-6331 (Print){\textbackslash}r0091-6331 (Linking)}
\showISSN{00916331}
\showDOI{%
\url{https://doi.org/10.1097/00003677-200104000-00009}}


\bibitem[\protect\citeauthoryear{D{\'{e}}gano, Marrugat, Grau,
  Salvador-Gonz{\'{a}}lez, Ramos, Zamora, Mart{\'{i}}, and Elosua}{D{\'{e}}gano
  et~al\mbox{.}}{2017}]%
        {Degano2017TheIndex}
\bibfield{author}{\bibinfo{person}{Irene~R. D{\'{e}}gano},
  \bibinfo{person}{Jaume Marrugat}, \bibinfo{person}{Maria Grau},
  \bibinfo{person}{Betlem Salvador-Gonz{\'{a}}lez}, \bibinfo{person}{Rafel
  Ramos}, \bibinfo{person}{Alberto Zamora}, \bibinfo{person}{Ruth Mart{\'{i}}},
  {and} \bibinfo{person}{Roberto Elosua}.} \bibinfo{year}{2017}\natexlab{}.
\newblock \showarticletitle{{The association between education and
  cardiovascular disease incidence is mediated by hypertension, diabetes, and
  body mass index}}.
\newblock \bibinfo{journal}{{\em Scientific Reports\/}} \bibinfo{volume}{7},
  \bibinfo{number}{1} (\bibinfo{year}{2017}), \bibinfo{pages}{1--8}.
\newblock
\showISBNx{2045-2322}
\showISSN{20452322}
\showDOI{%
\url{https://doi.org/10.1038/s41598-017-10775-3}}


\bibitem[\protect\citeauthoryear{Dorsey and Marks Jr.}{Dorsey and
  Marks Jr.}{2017}]%
        {Dorsey2017VerilyBiomarkers}
\bibfield{author}{\bibinfo{person}{E. Ray Dorsey} {and}
  \bibinfo{person}{William J. Marks Jr.}} \bibinfo{year}{2017}\natexlab{}.
\newblock \showarticletitle{{Verily and Its Approach to Digital Biomarkers}}.
\newblock \bibinfo{journal}{{\em Digital Biomarkers\/}}
  \bibinfo{volume}{94080} (\bibinfo{year}{2017}), \bibinfo{pages}{96--99}.
\newblock
\showISSN{2504-110X}
\showDOI{%
\url{https://doi.org/10.1159/000476051}}


\bibitem[\protect\citeauthoryear{Duan, Kravitz, and Schmid}{Duan
  et~al\mbox{.}}{2013}]%
        {Duan2013Single-patientResearch}
\bibfield{author}{\bibinfo{person}{Naihua Duan}, \bibinfo{person}{Richard~L.
  Kravitz}, {and} \bibinfo{person}{Christopher~H. Schmid}.}
  \bibinfo{year}{2013}\natexlab{}.
\newblock \showarticletitle{{Single-patient (n-of-1) trials: a pragmatic
  clinical decision methodology for patient-centered comparative effectiveness
  research}}.
\newblock \bibinfo{journal}{{\em Journal of Clinical Epidemiology\/}}
  \bibinfo{volume}{66}, \bibinfo{number}{8} (\bibinfo{date}{8}
  \bibinfo{year}{2013}), \bibinfo{pages}{S21--S28}.
\newblock
\showISSN{0895-4356}
\showDOI{%
\url{https://doi.org/10.1016/J.JCLINEPI.2013.04.006}}


\bibitem[\protect\citeauthoryear{Farseev and Chua}{Farseev and Chua}{2017}]%
        {Farseev2017TweetLearning}
\bibfield{author}{\bibinfo{person}{Aleksandr Farseev} {and}
  \bibinfo{person}{Tat-Seng Chua}.} \bibinfo{year}{2017}\natexlab{}.
\newblock \showarticletitle{{Tweet Can Be Fit: Integrating Data from Wearable
  Sensors and Multiple Social Networks for Wellness Profile Learning}}.
\newblock \bibinfo{journal}{{\em ACM Transactions on Information Systems\/}}
  \bibinfo{volume}{35}, \bibinfo{number}{4} (\bibinfo{date}{8}
  \bibinfo{year}{2017}), \bibinfo{pages}{1--34}.
\newblock
\showDOI{%
\url{https://doi.org/10.1145/3086676}}


\bibitem[\protect\citeauthoryear{{Firstbeat}}{{Firstbeat}}{2014}]%
        {Firstbeat2014AutomatedData}
\bibfield{author}{\bibinfo{person}{{Firstbeat}}.}
  \bibinfo{year}{2014}\natexlab{}.
\newblock \showarticletitle{{Automated Fitness Level ( VO 2 max ) Estimation
  with Heart Rate and Speed Data}}.
\newblock \bibinfo{journal}{{\em Firstbeat\/}} (\bibinfo{year}{2014}),
  \bibinfo{pages}{1--9}.
\newblock


\bibitem[\protect\citeauthoryear{Francis}{Francis}{2015}]%
        {Francis2015AMedicine}
\bibfield{author}{\bibinfo{person}{Collins Francis}.}
  \bibinfo{year}{2015}\natexlab{}.
\newblock \showarticletitle{{A New Initiative on Precision Medicine}}.
\newblock \bibinfo{journal}{{\em The New England Journal of Medicine\/}}
  \bibinfo{volume}{372}, \bibinfo{number}{9} (\bibinfo{year}{2015}),
  \bibinfo{pages}{1--3}.
\newblock
\showISBNx{0028-4793}
\showISSN{15334406}
\showDOI{%
\url{https://doi.org/10.1056/NEJMp1002530}}


\bibitem[\protect\citeauthoryear{Gabler, Duan, Vohra, and Kravitz}{Gabler
  et~al\mbox{.}}{2011}]%
        {Gabler2011N-of-1Review}
\bibfield{author}{\bibinfo{person}{Nicole~B. Gabler}, \bibinfo{person}{Naihua
  Duan}, \bibinfo{person}{Sunita Vohra}, {and} \bibinfo{person}{Richard~L.
  Kravitz}.} \bibinfo{year}{2011}\natexlab{}.
\newblock \bibinfo{title}{{N-of-1 Trials in the Medical Literature: A
  Systematic Review}}.
\newblock   (\bibinfo{year}{2011}), \bibinfo{numpages}{761--768}~pages.
\newblock
\showDOI{%
\url{https://doi.org/10.2307/23053842}}


\bibitem[\protect\citeauthoryear{Goff, Lloyd-Jones, Bennett, Coady, D'Agostino,
  Gibbons, Greenland, Lackland, Levy, O'Donnell, Robinson, Schwartz, Shero,
  Smith, Sorlie, Stone, and Wilson}{Goff et~al\mbox{.}}{2014}]%
        {Goff20142013Guidelines}
\bibfield{author}{\bibinfo{person}{David~C. Goff}, \bibinfo{person}{Donald~M.
  Lloyd-Jones}, \bibinfo{person}{Glen Bennett}, \bibinfo{person}{Sean Coady},
  \bibinfo{person}{Ralph~B. D'Agostino}, \bibinfo{person}{Raymond Gibbons},
  \bibinfo{person}{Philip Greenland}, \bibinfo{person}{Daniel~T. Lackland},
  \bibinfo{person}{Daniel Levy}, \bibinfo{person}{Christopher~J. O'Donnell},
  \bibinfo{person}{Jennifer~G. Robinson}, \bibinfo{person}{J.~Sanford
  Schwartz}, \bibinfo{person}{Susan~T. Shero}, \bibinfo{person}{Sidney~C.
  Smith}, \bibinfo{person}{Paul Sorlie}, \bibinfo{person}{Neil~J. Stone}, {and}
  \bibinfo{person}{Peter~W.F. Wilson}.} \bibinfo{year}{2014}\natexlab{}.
\newblock \showarticletitle{{2013 ACC/AHA guideline on the assessment of
  cardiovascular risk: A report of the American college of cardiology/American
  heart association task force on practice guidelines}}.
\newblock \bibinfo{journal}{{\em Circulation\/}} \bibinfo{volume}{129},
  \bibinfo{number}{25 SUPPL. 1} (\bibinfo{year}{2014}).
\newblock
\showISBNx{1524-4539 (Electronic){\textbackslash}r0009-7322 (Linking)}
\showISSN{15244539}
\showDOI{%
\url{https://doi.org/10.1161/01.cir.0000437741.48606.98}}


\bibitem[\protect\citeauthoryear{Hamilton, Hamilton, and Zderic}{Hamilton
  et~al\mbox{.}}{2007}]%
        {Hamilton2007RoleDisease}
\bibfield{author}{\bibinfo{person}{M~T Hamilton}, \bibinfo{person}{D~G
  Hamilton}, {and} \bibinfo{person}{T~W Zderic}.}
  \bibinfo{year}{2007}\natexlab{}.
\newblock \showarticletitle{{Role of low energy expenditure and sitting in
  obesity, metabolic syndrome, type 2 diabetes, and cardiovascular disease}}.
\newblock \bibinfo{journal}{{\em Diabetes\/}} \bibinfo{volume}{56},
  \bibinfo{number}{November} (\bibinfo{year}{2007}),
  \bibinfo{pages}{2655--2667}.
\newblock
\showISBNx{1939-327X (Electronic)}
\showISSN{1939-327X}
\showDOI{%
\url{https://doi.org/10.2337/db07-0882.CVD}}


\bibitem[\protect\citeauthoryear{Jain}{Jain}{2018}]%
        {Jain2018AHealth}
\bibfield{author}{\bibinfo{person}{Ramesh Jain}.}
  \bibinfo{year}{2018}\natexlab{}.
\newblock \showarticletitle{{A Navigational Approach to Health}}.
\newblock \bibinfo{journal}{{\em Arxiv\/}} (\bibinfo{date}{5}
  \bibinfo{year}{2018}).
\newblock
\showURL{%
\url{http://arxiv.org/abs/1805.05402}}


\bibitem[\protect\citeauthoryear{Jones and Carter}{Jones and Carter}{2000}]%
        {Jones2000TheFitness}
\bibfield{author}{\bibinfo{person}{Andrew~M. Jones} {and}
  \bibinfo{person}{Helen Carter}.} \bibinfo{year}{2000}\natexlab{}.
\newblock \showarticletitle{{The Effect of Endurance Training on Parameters of
  Aerobic Fitness}}.
\newblock \bibinfo{journal}{{\em Sports Medicine\/}} \bibinfo{volume}{29},
  \bibinfo{number}{6} (\bibinfo{year}{2000}), \bibinfo{pages}{373--386}.
\newblock
\showISSN{0112-1642}
\showDOI{%
\url{https://doi.org/10.2165/00007256-200029060-00001}}


\bibitem[\protect\citeauthoryear{Kubota, Heiss, Maclehose, Roetker, and
  Folsom}{Kubota et~al\mbox{.}}{2017}]%
        {Kubota2017AssociationStudy}
\bibfield{author}{\bibinfo{person}{Yasuhiko Kubota}, \bibinfo{person}{Gerardo
  Heiss}, \bibinfo{person}{Richard~F. Maclehose}, \bibinfo{person}{Nicholas~S.
  Roetker}, {and} \bibinfo{person}{Aaron~R. Folsom}.}
  \bibinfo{year}{2017}\natexlab{}.
\newblock \showarticletitle{{Association of educational attainment with
  lifetime risk of cardiovascular disease the atherosclerosis risk in
  communities study}}.
\newblock \bibinfo{journal}{{\em JAMA Internal Medicine\/}}
  \bibinfo{volume}{177}, \bibinfo{number}{8} (\bibinfo{year}{2017}),
  \bibinfo{pages}{1165--1172}.
\newblock
\showISSN{21686106}
\showDOI{%
\url{https://doi.org/10.1001/jamainternmed.2017.1877}}


\bibitem[\protect\citeauthoryear{Kumar, Abowd, Abraham, Absi, Hnat, Hossain,
  Ives, Kerr, Marlin, Murphy, Rehg, and Nahum-shani}{Kumar
  et~al\mbox{.}}{2017}]%
        {Kumar2017PervasiveMD2K}
\bibfield{author}{\bibinfo{person}{Santosh Kumar}, \bibinfo{person}{Gregory
  Abowd}, \bibinfo{person}{William~T Abraham}, \bibinfo{person}{Mustafa Absi},
  \bibinfo{person}{Timothy Hnat}, \bibinfo{person}{Syed~Monowar Hossain},
  \bibinfo{person}{Zachary Ives}, \bibinfo{person}{Jacqueline Kerr},
  \bibinfo{person}{Benjamin~M Marlin}, \bibinfo{person}{Susan Murphy},
  \bibinfo{person}{James~M Rehg}, {and} \bibinfo{person}{Inbal Nahum-shani}.}
  \bibinfo{year}{2017}\natexlab{}.
\newblock \showarticletitle{{Pervasive Health: Center of Excellence for Mobile
  Sensor Data-to-Knowledge (MD2K)}}.
\newblock \bibinfo{journal}{{\em Pervasive Computing, IEEE\/}}
  (\bibinfo{year}{2017}), \bibinfo{pages}{18--22}.
\newblock


\bibitem[\protect\citeauthoryear{Lefevre, Lewis, Perrett, and Penke}{Lefevre
  et~al\mbox{.}}{2013}]%
        {Lefevre2013TellingMen}
\bibfield{author}{\bibinfo{person}{Carmen~E. Lefevre}, \bibinfo{person}{Gary~J.
  Lewis}, \bibinfo{person}{David~I. Perrett}, {and} \bibinfo{person}{Lars
  Penke}.} \bibinfo{year}{2013}\natexlab{}.
\newblock \showarticletitle{{Telling facial metrics: Facial width is associated
  with testosterone levels in men}}.
\newblock \bibinfo{journal}{{\em Evolution and Human Behavior\/}}
  \bibinfo{volume}{34}, \bibinfo{number}{4} (\bibinfo{year}{2013}),
  \bibinfo{pages}{273--279}.
\newblock
\showISBNx{1090-5138}
\showISSN{10905138}
\showDOI{%
\url{https://doi.org/10.1016/j.evolhumbehav.2013.03.005}}


\bibitem[\protect\citeauthoryear{Lester, Hartung, Pina, Libby, Borriello, and
  Duncan}{Lester et~al\mbox{.}}{2009}]%
        {Lester2009ValidatedSensor}
\bibfield{author}{\bibinfo{person}{Jonathan Lester}, \bibinfo{person}{Carl
  Hartung}, \bibinfo{person}{Laura Pina}, \bibinfo{person}{Ryan Libby},
  \bibinfo{person}{Gaetano Borriello}, {and} \bibinfo{person}{Glen Duncan}.}
  \bibinfo{year}{2009}\natexlab{}.
\newblock \showarticletitle{{Validated caloric expenditure estimation using a
  single body-worn sensor}}.
\newblock \bibinfo{journal}{{\em Proceedings of the 11th international
  conference on Ubiquitous computing - Ubicomp '09\/}} (\bibinfo{year}{2009}),
  \bibinfo{pages}{225}.
\newblock
\showISBNx{9781605584317}
\showDOI{%
\url{https://doi.org/10.1145/1620545.1620579}}


\bibitem[\protect\citeauthoryear{Lucia, Santalla, P{\'{e}}rez, Chicharro, Luc,
  Hoyos, Rez, and Chicharro}{Lucia et~al\mbox{.}}{2002}]%
        {Lucia2002KineticsCyclists}
\bibfield{author}{\bibinfo{person}{Alejandro Lucia}, \bibinfo{person}{Alfredo
  Santalla}, \bibinfo{person}{Margarita P{\'{e}}rez},
  \bibinfo{person}{Luis~Miguel Chicharro}, \bibinfo{person}{Alejandro Luc},
  \bibinfo{person}{Jes~S Hoyos}, \bibinfo{person}{Margarita~Pé Rez}, {and}
  \bibinfo{person}{José~L Chicharro}.} \bibinfo{year}{2002}\natexlab{}.
\newblock \showarticletitle{{Kinetics of VO2 in professional cyclists}}.
\newblock \bibinfo{journal}{{\em Med. Sci. Sports Exerc\/}}
  \bibinfo{volume}{34}, \bibinfo{number}{2} (\bibinfo{year}{2002}),
  \bibinfo{pages}{320--325}.
\newblock
\showURL{%
\url{https://www.researchgate.net/publication/11532623}}


\bibitem[\protect\citeauthoryear{Maier, Schmid, M{\"{u}}ller, Steiner, and
  Wehrlin}{Maier et~al\mbox{.}}{2017}]%
        {Maier2017AccuracyCycling}
\bibfield{author}{\bibinfo{person}{Thomas Maier}, \bibinfo{person}{Lucas
  Schmid}, \bibinfo{person}{Beat M{\"{u}}ller}, \bibinfo{person}{Thomas
  Steiner}, {and} \bibinfo{person}{Jon Wehrlin}.}
  \bibinfo{year}{2017}\natexlab{}.
\newblock \showarticletitle{{Accuracy of Cycling Power Meters against a
  Mathematical Model of Treadmill Cycling}}.
\newblock \bibinfo{journal}{{\em International Journal of Sports Medicine\/}}
  \bibinfo{volume}{38}, \bibinfo{number}{06} (\bibinfo{date}{6}
  \bibinfo{year}{2017}), \bibinfo{pages}{456--461}.
\newblock
\showISSN{0172-4622}
\showDOI{%
\url{https://doi.org/10.1055/s-0043-102945}}


\bibitem[\protect\citeauthoryear{M{\"{u}}nzel, Schmidt, Steven, Herzog, Daiber,
  and S{\o}rensen}{M{\"{u}}nzel et~al\mbox{.}}{2018}]%
        {Munzel2018EnvironmentalSystem}
\bibfield{author}{\bibinfo{person}{Thomas M{\"{u}}nzel},
  \bibinfo{person}{Frank~P. Schmidt}, \bibinfo{person}{Sebastian Steven},
  \bibinfo{person}{Johannes Herzog}, \bibinfo{person}{Andreas Daiber}, {and}
  \bibinfo{person}{Mette S{\o}rensen}.} \bibinfo{year}{2018}\natexlab{}.
\newblock \showarticletitle{{Environmental Noise and the Cardiovascular
  System}}.
\newblock \bibinfo{journal}{{\em Journal of the American College of
  Cardiology\/}} \bibinfo{volume}{71}, \bibinfo{number}{6} (\bibinfo{date}{2}
  \bibinfo{year}{2018}), \bibinfo{pages}{688--697}.
\newblock
\showISSN{07351097}
\showDOI{%
\url{https://doi.org/10.1016/j.jacc.2017.12.015}}


\bibitem[\protect\citeauthoryear{Nag, Pandey, and Jain}{Nag
  et~al\mbox{.}}{2017a}]%
        {Nag2017b}
\bibfield{author}{\bibinfo{person}{Nitish Nag}, \bibinfo{person}{Vaibhav
  Pandey}, {and} \bibinfo{person}{Ramesh Jain}.}
  \bibinfo{year}{2017}\natexlab{a}.
\newblock \showarticletitle{{Health Multimedia: Lifestyle Recommendations Based
  on Diverse Observations}}.
\newblock \bibinfo{journal}{{\em Proceedings of the 2017 ACM on International
  Conference on Multimedia Retrieval\/}} (\bibinfo{year}{2017}),
  \bibinfo{pages}{99--106}.
\newblock
\showISBNx{978-1-4503-4701-3}
\showDOI{%
\url{https://doi.org/10.1145/3078971.3080545}}


\bibitem[\protect\citeauthoryear{Nag, Pandey, Oh, and Jain}{Nag
  et~al\mbox{.}}{2017b}]%
        {Nag2017a}
\bibfield{author}{\bibinfo{person}{Nitish Nag}, \bibinfo{person}{Vaibhav
  Pandey}, \bibinfo{person}{Hyungik Oh}, {and} \bibinfo{person}{Ramesh Jain}.}
  \bibinfo{year}{2017}\natexlab{b}.
\newblock \showarticletitle{{Cybernetic Health}}.
\newblock \bibinfo{journal}{{\em arXiv\/}} \bibinfo{volume}{arXiv:1705},
  \bibinfo{number}{May} (\bibinfo{date}{5} \bibinfo{year}{2017}).
\newblock
\showURL{%
\url{http://arxiv.org/abs/1705.08514}}


\bibitem[\protect\citeauthoryear{{Norbert Wiener}}{{Norbert Wiener}}{1948}]%
        {NorbertWiener1948Cybernetics:Machine}
\bibfield{author}{\bibinfo{person}{{Norbert Wiener}}.}
  \bibinfo{year}{1948}\natexlab{}.
\newblock \bibinfo{booktitle}{{\em {Cybernetics: Communication And Control In
  The Animal And The Machine}}}.
\newblock
\showURL{%
\url{https://archive.org/details/CyberneticsOrCommunicationAndControlInTheAnimalAndTheMachineNorbertWiener}}


\bibitem[\protect\citeauthoryear{{OpenCV}}{{OpenCV}}{[n. d.]}]%
        {OpenCVFacialDetection}
\bibfield{author}{\bibinfo{person}{{OpenCV}}.} \bibinfo{year}{[n.
  d.]}\natexlab{}.
\newblock \bibinfo{title}{{Facial Landmark Detection}}.
\newblock   (\bibinfo{year}{[n. d.]}).
\newblock
\showURL{%
\url{https://www.learnopencv.com/facial-landmark-detection/}}


\bibitem[\protect\citeauthoryear{Oskui, French, Herring, Mayeda, Burstein, and
  Kloner}{Oskui et~al\mbox{.}}{2013}]%
        {Oskui2013TestosteroneLiterature}
\bibfield{author}{\bibinfo{person}{P.~M. Oskui}, \bibinfo{person}{W.~J.
  French}, \bibinfo{person}{M.~J. Herring}, \bibinfo{person}{G.~S. Mayeda},
  \bibinfo{person}{S. Burstein}, {and} \bibinfo{person}{R.~A. Kloner}.}
  \bibinfo{year}{2013}\natexlab{}.
\newblock \showarticletitle{{Testosterone and the Cardiovascular System: A
  Comprehensive Review of the Clinical Literature}}.
\newblock \bibinfo{journal}{{\em Journal of the American Heart Association\/}}
  \bibinfo{volume}{2}, \bibinfo{number}{6} (\bibinfo{year}{2013}),
  \bibinfo{pages}{e000272--e000272}.
\newblock
\showISBNx{2047-9980 (Electronic){\textbackslash}r2047-9980 (Linking)}
\showISSN{2047-9980}
\showDOI{%
\url{https://doi.org/10.1161/JAHA.113.000272}}


\bibitem[\protect\citeauthoryear{P{\"{a}}rkk{\"{a}}, Ermes, Antila, Van~Gils,
  M{\"{a}}ntt{\"{a}}ri, and Nieminen}{P{\"{a}}rkk{\"{a}} et~al\mbox{.}}{2007}]%
        {Parkka2007EstimatingLocations}
\bibfield{author}{\bibinfo{person}{J. P{\"{a}}rkk{\"{a}}}, \bibinfo{person}{M.
  Ermes}, \bibinfo{person}{K. Antila}, \bibinfo{person}{M. Van~Gils},
  \bibinfo{person}{A. M{\"{a}}ntt{\"{a}}ri}, {and} \bibinfo{person}{H.
  Nieminen}.} \bibinfo{year}{2007}\natexlab{}.
\newblock \showarticletitle{{Estimating intensity of physical activity: A
  comparison of wearable accelerometer and gyro sensors and 3 sensor
  locations}}.
\newblock \bibinfo{journal}{{\em Annual International Conference of the IEEE
  Engineering in Medicine and Biology - Proceedings\/}} (\bibinfo{year}{2007}),
  \bibinfo{pages}{1511--1514}.
\newblock
\showISBNx{1424407885}
\showISSN{05891019}
\showDOI{%
\url{https://doi.org/10.1109/IEMBS.2007.4352588}}


\bibitem[\protect\citeauthoryear{Ross, Blair, Arena, Church, Despr{\'{e}}s,
  Franklin, Haskell, Kaminsky, Levine, Lavie, Myers, Niebauer, Sallis, Sawada,
  Sui, and Wisl{\o}ff}{Ross et~al\mbox{.}}{2016}]%
        {Ross2016ImportanceAssociation}
\bibfield{author}{\bibinfo{person}{Robert Ross}, \bibinfo{person}{Steven~N.
  Blair}, \bibinfo{person}{Ross Arena}, \bibinfo{person}{Timothy~S. Church},
  \bibinfo{person}{Jean~Pierre Despr{\'{e}}s}, \bibinfo{person}{Barry~A.
  Franklin}, \bibinfo{person}{William~L. Haskell}, \bibinfo{person}{Leonard~A.
  Kaminsky}, \bibinfo{person}{Benjamin~D. Levine}, \bibinfo{person}{Carl~J.
  Lavie}, \bibinfo{person}{Jonathan Myers}, \bibinfo{person}{Josef Niebauer},
  \bibinfo{person}{Robert Sallis}, \bibinfo{person}{Susumu~S. Sawada},
  \bibinfo{person}{Xuemei Sui}, {and} \bibinfo{person}{Ulrik Wisl{\o}ff}.}
  \bibinfo{year}{2016}\natexlab{}.
\newblock \bibinfo{booktitle}{{\em {Importance of Assessing Cardiorespiratory
  Fitness in Clinical Practice: A Case for Fitness as a Clinical Vital Sign: A
  Scientific Statement from the American Heart Association}}}.
  Vol.~\bibinfo{volume}{134}.
\newblock e653--e699 pages.
\newblock
\showISBNx{0000000000000}
\showISSN{15244539}
\showDOI{%
\url{https://doi.org/10.1161/CIR.0000000000000461}}


\bibitem[\protect\citeauthoryear{Sam~Gambhir, Jessie~Ge, Vermesh, and
  Spitler}{Sam~Gambhir et~al\mbox{.}}{2018}]%
        {SamGambhir2018TowardHealth}
\bibfield{author}{\bibinfo{person}{Sanjiv Sam~Gambhir}, \bibinfo{person}{T
  Jessie~Ge}, \bibinfo{person}{Ophir Vermesh}, {and} \bibinfo{person}{Ryan
  Spitler}.} \bibinfo{year}{2018}\natexlab{}.
\newblock \showarticletitle{{Toward achieving precision health}}.
\newblock \bibinfo{journal}{{\em Sci. Transl. Med\/}}  \bibinfo{volume}{10}
  (\bibinfo{year}{2018}).
\newblock
\showURL{%
\url{http://stm.sciencemag.org/content/scitransmed/10/430/eaao3612.full.pdf}}


\bibitem[\protect\citeauthoryear{Scheer, Hilton, Mantzoros, and Shea}{Scheer
  et~al\mbox{.}}{2009}]%
        {Scheer2009AdverseMisalignment}
\bibfield{author}{\bibinfo{person}{F.~A. J.~L. Scheer}, \bibinfo{person}{M.~F.
  Hilton}, \bibinfo{person}{C.~S. Mantzoros}, {and} \bibinfo{person}{S.~A.
  Shea}.} \bibinfo{year}{2009}\natexlab{}.
\newblock \showarticletitle{{Adverse metabolic and cardiovascular consequences
  of circadian misalignment}}.
\newblock \bibinfo{journal}{{\em Proceedings of the National Academy of
  Sciences\/}} \bibinfo{volume}{106}, \bibinfo{number}{11}
  (\bibinfo{year}{2009}), \bibinfo{pages}{4453--4458}.
\newblock
\showISBNx{1091-6490 (Electronic){\textbackslash}r0027-8424 (Linking)}
\showISSN{0027-8424}
\showDOI{%
\url{https://doi.org/10.1073/pnas.0808180106}}


\bibitem[\protect\citeauthoryear{Sedgwick}{Sedgwick}{2014}]%
        {Sedgwick2014WhatTrial}
\bibfield{author}{\bibinfo{person}{P. Sedgwick}.}
  \bibinfo{year}{2014}\natexlab{}.
\newblock \showarticletitle{{What is an {\&}quot;n-of-1{\&}quot; trial?}}
\newblock \bibinfo{journal}{{\em BMJ\/}} \bibinfo{volume}{348},
  \bibinfo{number}{apr10 1} (\bibinfo{date}{4} \bibinfo{year}{2014}),
  \bibinfo{pages}{g2674--g2674}.
\newblock
\showISSN{1756-1833}
\showDOI{%
\url{https://doi.org/10.1136/bmj.g2674}}


\bibitem[\protect\citeauthoryear{Senn}{Senn}{2017}]%
        {Senn2017SampleTrials}
\bibfield{author}{\bibinfo{person}{Stephen Senn}.}
  \bibinfo{year}{2017}\natexlab{}.
\newblock \showarticletitle{{Sample size considerations for <i>n</i> -of-1
  trials}}.
\newblock \bibinfo{journal}{{\em Statistical Methods in Medical Research\/}}
  (\bibinfo{date}{9} \bibinfo{year}{2017}), \bibinfo{pages}{096228021772680}.
\newblock
\showISSN{0962-2802}
\showDOI{%
\url{https://doi.org/10.1177/0962280217726801}}


\bibitem[\protect\citeauthoryear{Simon, Garg, Hunter, Guo, and Semega}{Simon
  et~al\mbox{.}}{2004}]%
        {Simon2004SensorEngines}
\bibfield{author}{\bibinfo{person}{D~L Simon}, \bibinfo{person}{S Garg},
  \bibinfo{person}{G~W Hunter}, \bibinfo{person}{T~H Guo}, {and}
  \bibinfo{person}{K~J Semega}.} \bibinfo{year}{2004}\natexlab{}.
\newblock \showarticletitle{{Sensor Needs for Control and Health Management of
  Intelligence Aircraft Engines}}.
\newblock  \bibinfo{number}{August} (\bibinfo{year}{2004}),
  \bibinfo{pages}{1--15}.
\newblock
\showISBNx{3016210134}
\showURL{%
\url{http://oai.dtic.mil/oai/oai?verb=getRecord&metadataPrefix=html&identifier=ADA426767}}


\bibitem[\protect\citeauthoryear{Smolander, Juuti, Kinnunen, Laine, Louhevaara,
  M{\"{a}}nnikk{\"{o}}, and Rusko}{Smolander et~al\mbox{.}}{2008}]%
        {Smolander2008AWorkers}
\bibfield{author}{\bibinfo{person}{Juhani Smolander}, \bibinfo{person}{Tanja
  Juuti}, \bibinfo{person}{Marja~Liisa Kinnunen}, \bibinfo{person}{Kari Laine},
  \bibinfo{person}{Veikko Louhevaara}, \bibinfo{person}{Kaisa
  M{\"{a}}nnikk{\"{o}}}, {and} \bibinfo{person}{Heikki Rusko}.}
  \bibinfo{year}{2008}\natexlab{}.
\newblock \showarticletitle{{A new heart rate variability-based method for the
  estimation of oxygen consumption without individual laboratory calibration:
  Application example on postal workers}}.
\newblock \bibinfo{journal}{{\em Applied Ergonomics\/}} \bibinfo{volume}{39},
  \bibinfo{number}{3} (\bibinfo{year}{2008}), \bibinfo{pages}{325--331}.
\newblock
\showISBNx{4589424894}
\showISSN{00036870}
\showDOI{%
\url{https://doi.org/10.1016/j.apergo.2007.09.001}}


\bibitem[\protect\citeauthoryear{Sun, Hong, and Wold}{Sun
  et~al\mbox{.}}{2010}]%
        {Sun2010CardiovascularExposure}
\bibfield{author}{\bibinfo{person}{Qinghua Sun}, \bibinfo{person}{Xinru Hong},
  {and} \bibinfo{person}{Loren~E. Wold}.} \bibinfo{year}{2010}\natexlab{}.
\newblock \showarticletitle{{Cardiovascular effects of ambient particulate air
  pollution exposure}}.
\newblock \bibinfo{journal}{{\em Circulation\/}} \bibinfo{volume}{121},
  \bibinfo{number}{25} (\bibinfo{year}{2010}), \bibinfo{pages}{2755--2765}.
\newblock
\showISBNx{1524-4539 (Electronic){\textbackslash}r0009-7322 (Linking)}
\showISSN{00097322}
\showDOI{%
\url{https://doi.org/10.1161/CIRCULATIONAHA.109.893461}}


\bibitem[\protect\citeauthoryear{Topol}{Topol}{2015}]%
        {topol-2}
\bibfield{author}{\bibinfo{person}{Eric Topol}.}
  \bibinfo{year}{2015}\natexlab{}.
\newblock \bibinfo{booktitle}{{\em {The patient will see you now: the future of
  medicine is in your hands}}}.
\newblock \bibinfo{publisher}{Basic Books}.
\newblock


\bibitem[\protect\citeauthoryear{{United States of America}}{{United States of
  America}}{[n. d.]a}]%
        {UnitedStatesofAmericaAirNowAgency}
\bibfield{author}{\bibinfo{person}{{United States of America}}.}
  \bibinfo{year}{[n. d.]}\natexlab{a}.
\newblock \bibinfo{title}{{AirNow - Environmental Protection Agency}}.
\newblock   (\bibinfo{year}{[n. d.]}).
\newblock
\showURL{%
\url{https://www.airnow.gov/}}


\bibitem[\protect\citeauthoryear{{United States of America}}{{United States of
  America}}{[n. d.]b}]%
        {UnitedStatesofAmericaCentersPrevention}
\bibfield{author}{\bibinfo{person}{{United States of America}}.}
  \bibinfo{year}{[n. d.]}\natexlab{b}.
\newblock \bibinfo{title}{{Centers for Disease Control and Prevention}}.
\newblock   (\bibinfo{year}{[n. d.]}).
\newblock
\showURL{%
\url{https://www.cdc.gov/}}


\bibitem[\protect\citeauthoryear{{United States of America - Depart of
  Transportation}}{{United States of America - Depart of Transportation}}{[n.
  d.]}]%
        {UnitedStatesofAmerica-DepartofTransportationNationalStatistics}
\bibfield{author}{\bibinfo{person}{{United States of America - Depart of
  Transportation}}.} \bibinfo{year}{[n. d.]}\natexlab{}.
\newblock \bibinfo{title}{{National Transportation Noise Map | Bureau of
  Transportation Statistics}}.
\newblock   (\bibinfo{year}{[n. d.]}).
\newblock
\showURL{%
\url{https://www.bts.gov/newsroom/national-transportation-noise-map}}


\bibitem[\protect\citeauthoryear{{United States of America - NOAA}}{{United
  States of America - NOAA}}{[n. d.]}]%
        {UnitedStatesofAmerica-NOAALightMap}
\bibfield{author}{\bibinfo{person}{{United States of America - NOAA}}.}
  \bibinfo{year}{[n. d.]}\natexlab{}.
\newblock \bibinfo{title}{{Light Pollution Map}}.
\newblock   (\bibinfo{year}{[n. d.]}).
\newblock
\showURL{%
\url{https://www.lightpollutionmap.info}}


\bibitem[\protect\citeauthoryear{Wild}{Wild}{2005}]%
        {Wild2005ComplementingEpidemiology}
\bibfield{author}{\bibinfo{person}{Christopher~Paul Wild}.}
  \bibinfo{year}{2005}\natexlab{}.
\newblock \showarticletitle{{Complementing the genome with an "exposome": The
  outstanding challenge of environmental exposure measurement in molecular
  epidemiology}}.
\newblock \bibinfo{journal}{{\em Cancer Epidemiology Biomarkers and
  Prevention\/}} \bibinfo{volume}{14}, \bibinfo{number}{8}
  (\bibinfo{year}{2005}), \bibinfo{pages}{1847--1850}.
\newblock
\showISBNx{1055-9965 (Print){\textbackslash}r1055-9965 (Linking)}
\showISSN{10559965}
\showDOI{%
\url{https://doi.org/10.1158/1055-9965.EPI-05-0456}}


\bibitem[\protect\citeauthoryear{Williams}{Williams}{2001}]%
        {Williams2001PhysicalMeta-Analysis}
\bibfield{author}{\bibinfo{person}{Paul~T Williams}.}
  \bibinfo{year}{2001}\natexlab{}.
\newblock \showarticletitle{{Physical fitness and activity as separate heart
  disease risk factors : a Meta-Analysis}}.
\newblock \bibinfo{journal}{{\em Med Sci Sports Exerc.\/}}
  \bibinfo{volume}{33}, \bibinfo{number}{5} (\bibinfo{year}{2001}),
  \bibinfo{pages}{754--761}.
\newblock


\bibitem[\protect\citeauthoryear{Wilson, D'Agostino, Sullivan, Parise, and
  Kannel}{Wilson et~al\mbox{.}}{2002}]%
        {Wilson2002OverweightRisk}
\bibfield{author}{\bibinfo{person}{Peter W.~F. Wilson},
  \bibinfo{person}{Ralph~B. D'Agostino}, \bibinfo{person}{Lisa Sullivan},
  \bibinfo{person}{Helen Parise}, {and} \bibinfo{person}{William~B. Kannel}.}
  \bibinfo{year}{2002}\natexlab{}.
\newblock \showarticletitle{{Overweight and Obesity as Determinants of
  Cardiovascular Risk}}.
\newblock \bibinfo{journal}{{\em Archives of Internal Medicine\/}}
  \bibinfo{volume}{162}, \bibinfo{number}{16} (\bibinfo{year}{2002}),
  \bibinfo{pages}{1867}.
\newblock
\showISBNx{0003-9926 (Print){\textbackslash}n0003-9926 (Linking)}
\showISSN{0003-9926}
\showDOI{%
\url{https://doi.org/10.1001/archinte.162.16.1867}}


\bibitem[\protect\citeauthoryear{Zhu, Pande, Mohapatra, and Han}{Zhu
  et~al\mbox{.}}{2015}]%
        {Zhu2015UsingSensors}
\bibfield{author}{\bibinfo{person}{Jindan Zhu}, \bibinfo{person}{Amit Pande},
  \bibinfo{person}{Prasant Mohapatra}, {and} \bibinfo{person}{Jay~J Han}.}
  \bibinfo{year}{2015}\natexlab{}.
\newblock \showarticletitle{{Using Deep Learning for Energy Expenditure
  Estimation with Wearable Sensors}}.
\newblock \bibinfo{journal}{{\em 17th International Conference on E-health
  Networking, Application {\&} Services (HealthCom) - IEEE\/}}
  (\bibinfo{year}{2015}), \bibinfo{pages}{501--506}.
\newblock
\showISBNx{9781467383257}
\showDOI{%
\url{https://doi.org/10.1109/HealthCom.2015.7454554}}


\end{thebibliography}
% \bibliography{sample-sigconf}

\end{document}